\RequirePackage{fix-cm}

\documentclass[onecolumn,epjc3]{svjour3}  
\usepackage{textcomp}
\usepackage{amssymb}
\usepackage{xcolor}
\smartqed  % flush right qed marks, e.g. at the end of the proof
\usepackage{amsmath}
\usepackage{graphicx}
\usepackage{subcaption}
\usepackage[a4paper, margin=2cm]{geometry}  % 可根据需求调整页边距

\journalname{Eur. Phys. J. C}
\begin{document}

\title{Revisiting the equation of state of dark energy from DESI BAO with SNe Ia and CMB}
% Identifying the Redshift Origin of DESI BAO Deviation from the $\Lambda$CDM model with CMB and SNe Ia
%\subtitle{Do you have a subtitle?\\ If so, write it here}

%\titlerunning{Short form of title}        % if too long for running head

\author{
Jie Zheng\thanksref{addr1}
\and
Da-Chun Qiang\thanksref{e2,addr2}
\and
Zhi-Qiang You\thanksref{addr2}
\and
Darshan Kumar\thanksref{addr2}
}

\thankstext{e2}{Corresponding author. E-mail: dcqiang@hnas.ac.cn}

\institute{
School of Microelectronics,
Qingdao University of Science and Technology,
Qingdao, China\label{addr1}
\and
Institute for Gravitational Wave Astronomy,
Henan Academy of Sciences,
Zhengzhou, China
\label{addr2}
}
\date{Received: date / Accepted: date}
% The correct dates will be entered by the editor

\maketitle

\begin{abstract}

The Dark Energy Spectroscopic Instrument (DESI) measurements of baryon acoustic oscillations (BAO) have recently shown a mild preference for dynamical dark energy over the standard $\rm \Lambda$CDM model.
In this paper, we analyze the $w_0w_a$CDM model using DESI BAO DR2, Pantheon+ SNe Ia, and Planck 2018 CMB distance prior data. To examine how different parts of the data affect the apparent deviation from $\rm \Lambda$CDM, we adopt two complementary redshift-cut strategies, dividing the dataset into $z<z_{\rm cut}$ and $z>z_{\rm cut}$ subsamples. We find that the most noticeable shifts occur when BAO and SNe Ia data in the redshift range $z\sim0.4$--$0.8$ are included, reaching a significance of about $\sim2\sigma$. Within this framework, the inclusion of higher-redshift measurements progressively weakens this deviation and brings the constraints closer to the $\rm \Lambda$CDM expectation. 
Moreover, the information criteria show no statistically significant preference between the $w_0w_a$CDM and $\rm \Lambda$CDM models, while the Bayesian information criterion consistently favors the $\rm \Lambda$CDM model. In addition, a parameter-shift consistency test reveals no statistically significant tension between complementary redshift subsamples.  Within the DESI BAO DR2, Pantheon+, and CMB distance prior framework adopted here, these results do not provide statistically robust evidence favoring the $w_0w_a$CDM model over $\rm \Lambda$CDM. They instead indicate that the apparent parameter shifts under different redshift cuts may be affected by statistical fluctuations and by the limited precision and number of datapoints in the current subsamples. Our analysis provides a complementary redshift-dependent diagnostic for assessing how the inferred cosmological constraints vary under different redshift selections.
%\keywords{Cosmological parameters \and dark energy \and quasars}
% \PACS{PACS code1 \and PACS code2 \and more}
% \subclass{MSC code1 \and MSC code2 \and more}
\end{abstract}

\section{Introduction}
\label{intro}
The $\rm \Lambda$CDM model has been tested and validated by a wide range of cosmological observations under the assumption that the Universe is spatially flat, isotropic, and homogeneous, with a constant $\rm \Lambda$ describing dark energy. However, in the past decade, some astronomical observations have revealed tensions in certain cosmological parameters within the $\rm \Lambda$CDM model \cite{2012MNRAS.427..146H,2018AAS...23231902P,2020MNRAS.498.1420W,2020NatAs...4..196D,2021A&A...646A.140H,Cao:2021zpf,2021PhRvD.103d1301H,2021ApJ...908L...9D,2022PhRvD.105b3520A}. 
For instance, the well-known $H_0$ tension refers to the discrepancy between the Hubble constant inferred from early-universe measurements of the cosmic microwave background (CMB) by Planck under the $\rm \Lambda$CDM model and that derived from late-universe observations of Cepheid variables and Type Ia supernovae (SNe Ia). In addition, recent baryon acoustic oscillation (BAO) measurements may hint at dynamical dark energy rather than a cosmological constant \cite{2025JCAP...04..012A}. 
The DESI collaboration reports that when combining their BAO Data Release 1 (DESI DR1) with Planck CMB data and different Type Ia supernova samples, namely Pantheonplus~\cite{pantheon+_1,pantheon+_2}, Union3~\cite{union3}, and the Dark Energy Survey Year-Five (DESY5) dataset~\cite{DESY5-1,DESY5-2,DESY5-3}, the significance levels of the deviation from the $\rm \Lambda$CDM model are $2.5\sigma$, $3.5\sigma$, and $3.9\sigma$, respectively~\cite{desidr1}. These discrepancies increase to $2.8\sigma$, $3.8\sigma$, and $4.2\sigma$ when the BAO data are updated from DESI DR1 to Data Release 2 (DR2)~\cite{DESI:2025zgx}. 
{The origin of these apparent deviations may arise from statistical fluctuations, unidentified systematic effects in the observational data, inconsistencies among different datasets, possible redshift-dependent effects in the measurements, or new physics beyond the standard $\rm \Lambda$CDM model. To clarify these issues, numerous studies have investigated the impact of different data combinations, subsample selections, and potential tensions among datasets, in addition to exploring alternative cosmological models (see e.g. Refs.~\cite{2025arXiv250204212H,2025arXiv250709981Q,William-CMB_cpl,huang_H0_CMB,2025SCPMA..6800413H,2025MNRAS.538..875E,2024JCAP...12..007C,2025arXiv250103480P,2025MNRAS.540.1626D,2024ApJ...976L..11R,2025JCAP...03..023A,2025arXiv250108915S,2025arXiv250114366S,2024arXiv241013627P,2024arXiv241212905C,2025EPJC...85..286O,2024arXiv240704385L,2024MNRAS.534.3869W,2024PhRvD.110l3512H,2024arXiv240408633C,2025PhRvD.112f3551G,2025arXiv250621010N,2026JHEAp..5000471O,2025JHEAp..4700398O,2025arXiv250416868A,2025ApJ...986L..31R,2024ApJ...976....1L,2025SCPMA..6910413L,2025JCAP...01..153J,2024PhRvD.110l3519J,2024PhRvD.110h3528W,2025PhRvD.111d1303W,2025JCAP...05..034W,Zheng_JCAP_CDDR,2025JCAP...08..056Z}).}

In particular, several works have investigated the role of specific observational probes in driving the reported deviation from the $\rm \Lambda$CDM model. 
For example, in BAO measurements, the LRG1 and LRG2 samples from DESI have been identified as key contributors to the apparent preference for dynamical dark energy~\cite{2024arXiv240704385L,2024MNRAS.534.3869W,2024PhRvD.110l3512H,2024arXiv240408633C}. 
These subsamples have also been reported to exhibit some tension with earlier BAO measurements from the Sloan Digital Sky Survey (SDSS), suggesting that the inferred deviation may be sensitive to specific BAO components.
In the case of CMB data, several studies have examined the impact of different CMB datasets on the deviation from $\rm \Lambda$CDM when combined with DESI BAO and SNe Ia. 
It has been shown that replacing or extending the Planck dataset with alternative CMB measurements can mitigate the observed deviation~\cite{2025PhRvD.112b3508G}. 
In addition, using CMB data without Planck leads to tighter constraints on the Hubble constant and reduces the significance of the Hubble tension~\cite{huang_H0_CMB}. 
Furthermore, lensing measurements from Planck and ACT-DR6 have also been found to contribute to the apparent indication of dynamical dark energy~\cite{2025arXiv250709981Q}. 
For SNe Ia datasets, differences among various compilations, calibration treatments, and subsample selections have been reported in the literature. In particular, analyses including the DESY5 sample have shown shifts in the inferred cosmological parameters compared with other datasets. These shifts may be related to several factors, including redshift coverage, sample composition, light-curve calibration, and selection effects~\cite{DESY5-1,DESY5-2,DESY5-3,2017AJ....154..211K,2009ApJ...700..331H,2012ApJS..200...12H,2018MNRAS.475..193F}.
Offsets between low- and high-redshift subsets have also been discussed for both DESY5 and Pantheon+ samples~\cite{2025MNRAS.538..875E,pantheon+_1,pantheon+_2}. In addition, different treatments of low-redshift SNe Ia, calibration systematics, and sample selections can change the inferred level of preference for dynamical dark energy~\cite{2025arXiv250204212H,2025arXiv250709981Q}.
Redshift-dependent analyses of supernova data have further revealed variations in inferred cosmological parameters under different redshift selections~\cite{2023ChPhC..47l5101T,2023PhRvD.108l3533M}.
These results do not necessarily indicate a single issue in any individual dataset, but they highlight that cosmological constraints from SNe Ia can be sensitive to calibration choices, redshift coverage, and subsample selection. This motivates a systematic investigation of the stability of dark energy constraints under different redshift cuts. Since our analysis adopts Pantheon+ as the only SNe Ia compilation and does not perform a systematic comparison with DESY5, Union3, or other samples, our conclusions should be interpreted within the Pantheon+ framework adopted here.

Most previous studies have focused on specific subsamples or comparisons between a limited number of selected redshift ranges. 
In contrast, we adopt a complementary redshift cutoff approach, dividing the dataset into two redshift subsamples, $z<z_{\rm cut}$ and $z>z_{\rm cut}$, and systematically comparing their inferred cosmological constraints within a unified statistical framework. This approach allows us to examine how the inferred cosmological constraints vary under different redshift selections and to assess the robustness of the apparent deviation from the $\rm \Lambda$CDM model. Compared to previous studies focusing on specific subsamples or individual redshift intervals, our approach provides a systematic and internally consistent way to assess the stability of the inferred constraints on the dark energy equation-of-state parameters under complementary redshift cuts.
In this paper, we perform a comprehensive analysis within the framework of the dynamical dark energy $w_0w_a$CDM model, where the equation-of-state parameters $w_0$ and $w_a$ are constrained using DESI BAO DR2, Pantheon+ SNe Ia, and CMB distance prior information. Our goal is not to attribute the deviation to specific redshift intervals, but to test the stability of the inferred cosmological constraints under different redshift selections.
To provide a quantitative assessment, we employ two statistical diagnostics.
First, information criteria are used to evaluate the relative preference of the dynamical dark energy $w_0w_a$CDM model compared with the $\rm \Lambda$CDM model. 
Second, a parameter-difference statistic, following the approach used in Planck analyses \cite{planck2018}, is applied to the complementary redshift subsamples defined by a threshold $z_{\rm cut}$ to quantify their statistical consistency and compute the associated probability-to-exceed (PTE). 
This test also enables us to assess whether any apparent parameter shifts arise from possible new physics or simply from the reduced amount of data after the redshift cut. The remainder of this paper is organized as follows. Section~\ref{sec:dam} describes the data and methodology. Section~\ref{sec:res} presents our results, and conclusions are drawn in Section~\ref{sec:con}.

\section{Data and methodoloy}
\label{sec:dam}
In this section, we first provide a brief introduction to the cosmological model before describing the datasets we used. We adopt the $w_{0}w_{a}$CDM model in our analysis. In this model, the behavior of the dark energy component varies with the redshift $z$, which can be written as
\begin{equation}
    w(z)=w_{0}+\frac{w_{a}z}{1+z}.
\end{equation}
And the corresponding Friedmann equation can be written as
\begin{equation}
\begin{split}
    E^{2}(z) = &\ \Omega_{m}(1+z)^3 + {\rm \Omega}_{r}(1+z)^{4}  \\
    & + {\rm \Omega}_{\mathrm{DE}}(1+z)^{3(1+w_{0}+w_{a})} 
    e^{-\frac{3 w_{a} z}{1+z}},
\end{split}
\end{equation}
where ${\rm \Omega}_{m}$, ${\rm \Omega}_{r}$, and ${\rm \Omega}_{\mathrm{DE}}$ are the present-time energy density of dark matter, radiation, and dark energy, respectively.

For our observational dataset, we use BAO measurements in galaxy, quasar, and Lyman-$\alpha$ forest tracers from the Year 1 data release from DESI, listed in Table.~\ref{tab:baodata}, the CMB acoustic peak positions from Planck 2018 \cite{planck2018}, and the latest SNe Ia compilation, Pantheonplus compilation \cite{pantheonp}.
%In order to clearly identify which redshift interval’s data might lead to deviations in the dark energy equation of state, each interval contains only one additional BAO data point.
We subdivide the BAO and SNe Ia samples into six different redshift cuts, as shown in Table.~\ref{tab:cut}, ensuring that each interval contains one additional BAO data point. Within each redshift cut, we measure the equation of state (EoS) parameters of dark energy, \( w_{0} \) and \( w_{a} \).

\subsection{Baryon Acoustic Oscillation (BAO)}

\begin{table*}
\centering
    \caption{The clustering measurement of DESI DR2 galaxies samples with distance ratios, including $D_{V}/r_{d}$, $D_{M}/D_{H}$, $D_{M}/r_{d}$, and $D_{H}/r_{d}$, at effective redshifts $z_{\mathrm{eff}}$, which is reproduced from Ref.~\cite{2024arXiv240403002D}. Also, the cross-correlation coefficients $r_{V,M/H}$ between $D_{V}/r_{d}$ and $D_{M}/D_{H}$, and $r_{M,H}$ between $D_{M}/r_{d}$ and $D_{M}/r_{d}$.}
	\centering
	\begin{tabular}{cccccccc} 
		\hline
		Tracer & $z_{\mathrm{eff}}$ & $D_{V}/r_{d}$ & $D_{M}/D_{H} $ & $r_{V,M/H}$ & $D_{M}/r_{d}$ & $D_{H}/r_{d}$ & $r_{M,H}$ \\
		\hline
		BGS & 0.295 & $7.944\pm0.075$ & - & - & - & - & - \\
            LRG1 & 0.510 & $12.720\pm0.098$ & $0.622\pm0.017$ & 0.064 & $17.347\pm0.180$ & $21.863\pm0.427$ & -0.475\\
            LRG2 & 0.706 & $16.048\pm0.110$ & $0.892\pm0.021$ & -0.001 & $17.347\pm0.180$ & $19.458\pm0.332$ & -0.423 \\
            LRG3+ELG1 & 0.934 & $19.720\pm0.091$& $1.223\pm0.019$& 0.067 & $21.574\pm0.153$ & $17.641\pm0.193$ & -0.425 \\
            ELG2 & 1.321 & $24.256\pm0.174$ & $1.948\pm0.044$ & 0.228 &  $27.605\pm0.320$ & $14.178\pm0.217$ & -0.437 \\
            QSO & 1.484 & $26.059\pm0.400$ & $2.386\pm0.135$ & 0.042 &  $30.519\pm0.758$ & $12.816\pm0.513$ & -0.489 \\
            Ly$\alpha$ QSO & 2.330 & $31.267\pm0.256$ & $4.518\pm0.097$ & 0.574 & $38.988\pm0.531$ & $8.632\pm0.101$ & -0.431 \\
		\hline
	\end{tabular}
    \label{tab:baodata}
\end{table*}

%\begin{figure*}
%    \centering
%    \includegraphics[width=\textwidth]{redshift_distribution.pdf}
%    \caption{Redshift cut}
%    \label{fig:distribution}
%\end{figure*}

\begin{table*}[htbp]
\centering
\caption{Two redshift cutoff approaches adopted for the DESI BAO and SNe Ia Pantheonplus datasets, including low-redshift path ($z<z_{\rm cut}$) and high-redshift path ($z>z_{\rm cut}$) for different values of $z_{\rm cut}$. It should be noted that the CMB dataset is excluded from this cutoff method. }
\label{tab:cut}
\begin{tabular}{lll}
\hline
\textbf{Cut Type} & \textbf{Redshift Range}                 & \textbf{Applied Datasets} \\
\hline
Low-redshift cut    & $z < 0.4$, $z<0.6$, $z<0.8$, $z< 1.0$, $z< 1.4$, $z < 1.6$ & BAO, SN Ia\\
High-redshift cut  & $z > 0.4$, $z >0.6$, $z >0.8$, $z >1.0$, $z >1.4$, $z >1.6$ & BAO, SN Ia \\
\hline
\end{tabular}
\end{table*}

The BAO measurements have been a powerful cosmological probe for investigating the physics of accleration, since the first clear detections of BAO in the Sloan Digital Sky Survey and the Two-Degree Field Galaxy Redshift Survey \cite{eisenstein1998,2001MNRAS.327.1297P,2005MNRAS.362..505C,eisenstein20005,2016RPPh...79d6902K,2016ARNPS..66...95J,2019LRR....22....1I,2021JCAP...11..050A}. The DESI collaboration is carrying out a State IV survey aimed at significantly enhancing cosmological constraints and it is conducting a 5-year survey that covers 1,4200 square degrees in the redshift range $0.1<z<4.2$ over five years, and its spectroscopic sample size is ten times larger than previous surveys \cite{2024arXiv240403000D}.
%Compared to the DR1, the DR2 are 2.4 times and 2.3 times the number of dark and bright tiles. 
In the DESI survey, the observational strategy is divided into two main programs, `bright' and `dark', that are categorized based on night-sky conditions, each of them focus on different sets of target classes \cite{2023AJ....165...50M,2023AJ....166..259S,2023AJ....165..253H,2023AJ....165...58Z}. The DR2 release covers 8 different classes of tracers. Under bright conditions, it includes the bright galaxy sample BGS ($0.1<z<0.4$). During dark conditions, it has luminous red galaxies LRG ($0.4<z<0.6$ and $0.6<z<0.8$), emission line galaxies ELG ($1.1<z<1.6$), the combined LRG and ELG sample (LRG+ELG, $0.8<z<1.1$), and quasars (QSO, $0.8<z<2.1$). Additionally, for the Lyman-$\alpha$ forest quasars (Ly$\alpha$ QSO, $1.77<z<4.16$), the survey measures the auto-correlation of the Ly$\alpha$ forest absorption in quasar spectra and the cross-correlation between the forest absorption and quasar positions. 

In the case of the flat universe, the BAO scale measurement in the transverse direction at redshift $z$ constraints the transverse comoving distance $D_M(z)$, which is given by
\begin{equation}
    \label{eq:DM}
    D_{M}=\frac{c}{H_{0}} \int_{0}^{z} \frac{d z^{\prime}}{E\left(z^{\prime}\right)},
\end{equation}
where $c$ is the velocity of light and $E(z)=H(z)/H_{0}$. 
The line-of-sight direction measurement constrains the expansion rate $H(z)$ or the corresponding distance $D_{H}(z)$, which is expressed as
\begin{equation}
\label{eq:DH}
D_{H}=\frac{c}{H(z)}.
\end{equation}
Besides, the isotropic BAO distance $D_{V}(z)$ is defined by
\begin{equation}
\label{eq:DV}
    D_{V}(z)=[zD_{M}^{2}(z)D_{H}(z)]^{1/3},
\end{equation}
The ratios $D_{M}/r_{d}$ and $D_{H}/r_{d}$ can be directly constrained, since the inferred distances are measured relative to the sound horizon $r_{d}$,
\begin{equation}
    \label{eq:rd}
    r_{d} = \int_{z_{d}}^{\infty} \frac{c_{s}(z)}{H(z)} dz,
\end{equation}
where $c_{s}(z)$ is the speed of sound in the photon-baryon fluid, and $z_{d}$ is the redshift at which acoustic waves stall because photons no longer `drag' the baryons. Here,  we do not calibrate the sound horizon parameter $r_{d}$, and we sample over the parameter $\mathrm{h} r_{d}$, where $\mathrm{h}\equiv H_{0}/(100\mathrm{km\ s^{-1}Mpc^{-1}})$. 
Then, the $\chi^{2}$ likelihood function of BAO data is expressed as
\begin{equation}
\chi^{2}=\Delta \vec{X}^{T} \cdot \mathbf{C}^{-1} \cdot \Delta \vec{X},
\end{equation}
where $\Delta \vec{X}=x_{i}-x_\mathrm{model}$, with $x_\mathrm{model}$ calculated by Eq.~\ref{eq:DM}-\ref{eq:rd} at the effective redshift $z_{\mathrm{eff}}$, and $x_{i}$ taken from Table.~\ref{tab:baodata}. 
Moreover, the covariance matrix \(\mathbf{C}\) can be determined based on the cross-correlation coefficients and the observational errors.

\subsection{Cosmic Microwave Background (CMB)}

The precise measurements of the CMB temperature and polarization anisotropies offer critical insights into the early universe and the formation of large-scale structures. As a fundamental cosmological probe, the CMB observations contain an abundance of information on key cosmological parameters \cite{2011ApJS..192...18K,2013ApJS..208...19H,2014A&A...571A..16P,planck2018}, including the Hubble constant and the properties of dark energy. 
{ Instead of using the full power of CMB information, we adopt the distance priors as in \cite{wang2007prd_dp,2019JCAP...02..028C}. This choice is motivated by the need to efficiently explore multiple redshift-cut configurations within a unified analysis framework. The distance priors compress the dominant geometric information of the CMB power spectra into a reduced set of parameters, namely the acoustic scale $l_{A}$, the ``shift parameter" $R$, and the redshift of recombination $z_{*}$. These quantities encapsulate the primary sensitivity of the CMB to the background expansion history. Since our primary goal is to investigate the relative impact of redshift cuts under a consistent statistical framework, this approach is adequate for our study.}
%This approach significantly reduces the computational cost while retaining the primary sensitivity of CMB observations to late-time background evolution. Although the use of distance priors may lead to slight shifts in the central values and uncertainties compared to a full-likelihood analysis, it preserves the essential degeneracy structure relevant for dark energy constraints. Since the purpose of this work is to investigate the relative impact of redshift cuts within a consistent analysis framework rather than to reproduce the exact numerical significance reported by the full Planck likelihood, the distance-prior approach is sufficient and appropriate for our analysis.

The definition of the distance priors is
\begin{equation}
    \label{eq:la}
    l_{A}=(1+z_{*})\frac{\pi D_{A}(z_{*})}{r_{s}(z_{*})},
\end{equation}
\begin{equation}
    \label{eq:R}
    R(z_{*})\equiv \frac{(1+z_{*})D_{A}(z_{*})\sqrt{{\rm \Omega}_{m}H_{0}}}{c},
\end{equation}
where $z_{*}$ is the redshift at the photon decoupling epoch, which can be approximately calculated by
\begin{equation}
    \label{eq:z*}
    z_{*}=1048[1+0.00124({\rm \Omega}_{b}h^{2})^{-0.738}][1+g_{1}({\rm \Omega}_{b}h^{2})^{g_{2}}],
\end{equation}
where 
\begin{equation}
    g_{1}=\frac{0.0738({\rm \Omega}_{b}h^{2})^{-0.238}}{1+39.5({\rm \Omega}_{b}h^{2})^{0.763}},
\end{equation}
\begin{equation}
    g_{2}=\frac{0.560}{1+21.1({\rm \Omega}_{b}h^{2})^{1.81}}.
\end{equation}
In Eq.~\ref{eq:la}, the comving sound horzion is defined by
\begin{equation}
    \label{eq:rs}
    r_{s}(z)=\frac{c}{H_{0}}\int^{a}_{0} \frac{da}{a^{2}E(a)\sqrt{3(1+\frac{3{\rm \Omega}_{b}h^{2}}{4{\rm \Omega}_{\gamma}h^{2}}a)}},
\end{equation}
where $a=1//(1+z)$, $\frac{3}{4{\rm \Omega}_{\gamma}h^{2}}=31500(T_{CMB}/2.7K)^{-4}$ and $T_{CMB}=2.755K$.
And the angular diameter distance $D_{A}$ is written by
\begin{equation}
    \label{eq:DA}
    D_{A}=\frac{c}{H_{0}}\int^{z}_{0} \frac{dz^{\prime}}{E(z^{\prime})}.
\end{equation}

The $\chi^{2}$ of the CMB is given by
\begin{equation}
    \chi^{2}=\Delta \vec{Y}^{T} \mathbf{C}^{-1} \Delta \vec{Y},
\end{equation}
where $\Delta \vec Y=y_{i}-y_{\mathrm{model}}$, with $y_{i}=\{1.7502, 301.471, 0.02236\}$ and $y_{\mathrm{model}}=\{R(z^{*}, l_{A}(z_{*}), {\rm \Omega}_{b}h^{2})\}$. Their correlation matrix $\mathbf{C}$ can be found in Ref.~\cite{2019JCAP...02..028C}.

\subsection{Type Ia supernovae (SNe Ia)}

Type Ia SNe are luminous standardizable candles that play a pivotal role in cosmology, particularly in studying the expansion history of the universe. Their well-characterized light curves and consistent peak luminosities enable precise distance measurements, making them invaluable tools for constraining cosmological parameters. Type Ia SNe are especially effective at low redshifts (\(0.01 < z < 0.3\)), where other cosmological probes, such as BAO, are limited by cosmic variance. In this work, we adopt the recent Pantheonplus compilation \cite{pantheonp}, which contains 1701 light curves of 1550 distinct SNe Ia spanning the redshift range $0.00122<z<2.26137$.
%This compilation is an extension of the Pantheon compilation \cite{pantheon}, with enhancements in sample size and improvements in the treatment of systematic uncertainties related to redshifts, peculiar velocities, photometric calibrations, and intrinsic scatter models of SNe Ia.
Here, we remove data points at $z < 0.01$ to avoid biases arising from the significant impact of peculiar velocities at very low redshifts $z < 0.008$ \cite{pantheonp}.

For the Pantheonplus compilation, the observed distance modulus is given by
\begin{equation}
\mu=m_{B}+\alpha x_{1}-\beta c-M-\delta_{\mathrm{bias}}+\delta_{\mathrm{host}},
\end{equation}
where $m_{B}$ denotes the apparent B-band magnitude, $x_{1}$ is the stretch parameter corresponding to light-curve width, $c$ is the light-curve color that includes contributions from both intrinsic color and dust, $M$ is the absolute B-band magnitude of a fiducial SNe Ia, $\delta_{\mathrm{bias}}$ is a correction term to account for selection biases, and $\delta_{\mathrm{host}}$ is the luminosity correction for residual correlations. $\alpha$ and $\beta$ are global nuisance parameters that relate stretch and color to luminosity, respectively. 
The theoretical distance modulus $\mu_{\mathrm{th}}$ is defined as 
\begin{equation}
    \mu_{\mathrm{th}}= 5\mathrm{log}_{10}\frac{D_{L}(z)}{\mathrm{Mpc}}+25,
\end{equation}
where $D_{\mathrm{L}}(z)$ is the luminosity distance, 
\begin{equation}
\label{eq:DL}
   D_L(z)=\frac{c(1+z)}{H_0}\int^{z}_{0} \frac{d \tilde{z}}{E(\tilde{z})}.
\end{equation}

Then, the $\chi^{2}$ likelihood function of Pantheonplus compliation can be written as
\begin{equation}
\chi^{2}=\Delta \vec{D}^{T} \cdot \mathbf{C}^{-1} \cdot \Delta \vec{D},
\end{equation}
where the covariance matrix $\mathbf{C}$ is the covariance matrix including both the systematic and statistical errors, which can be found on the website\footnote{https://github.com/PantheonPlusSH0ES/DataRelease}, and $\Delta \vec{D}$ is the vector of SNe Ia distance modulus residuals,
\begin{equation}
    \Delta \vec{D}_{i}=\mu_{i}-\mu_{\mathrm{model}}(z_{i}).
\end{equation}

%In the cosmological analysis, the probability distributions of model parameters are obtained with an affine invariant Markov chain Monte Carlo (MCMC) ensemble sampler (emcee) \cite{emcee}, where the statistic can be determined with
%\begin{equation}
%\mathcal{L}(p)=e^{-\frac{\chi(p)^{2}}{2}},
%\end{equation}
%where $p$ is the set of model parameters from different cosmological models. 

\section{Results and Discussion}
\label{sec:res}
\begin{figure*}
    \centering
    \includegraphics[width=\textwidth]{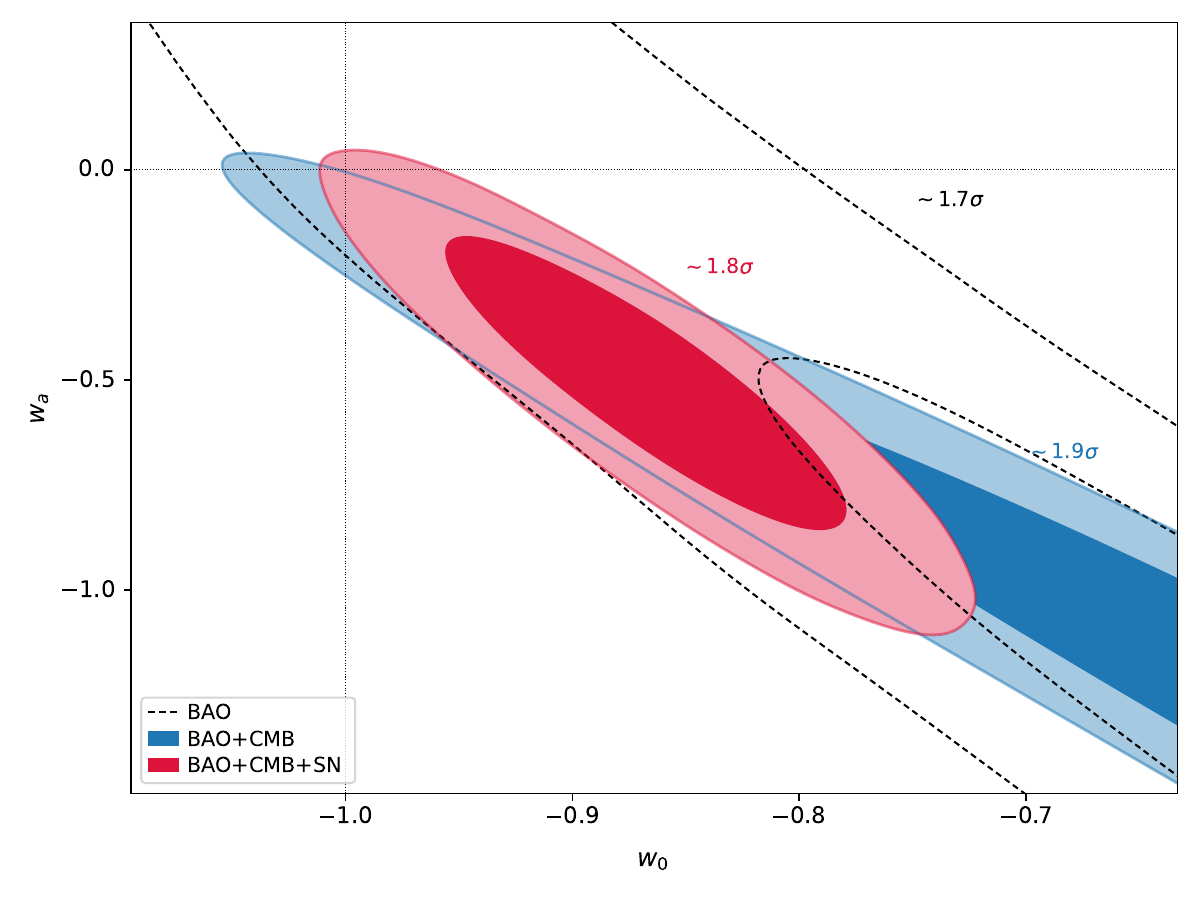}
    \caption{The posterior distributions of $(w_0, w_a)$ from the $w_0w_a$CDM fits to BAO (black dashed), BAO+CMB (blue), and BAO+CMB+SN (red), respectively. These contours indicate the 68\% and 95\% confidence levels. The gray dashed lines mark $w_0=-1$ and $w_a=0$ corresponding to the $\rm \Lambda$CDM model. The deviation from the $\rm \Lambda$CDM model is quantified at approximately $1.7\sigma$ for BAO, $1.9\sigma$ for BAO+CMB, and $1.8\sigma$ for BAO+CMB+SN.}
    \label{fig:all}
\end{figure*}

\begin{figure*}[p]  % [t] 也可以
    \centering
    \begin{subfigure}[t]{0.48\textwidth}
        \includegraphics[width=\linewidth,height=0.27\textheight]{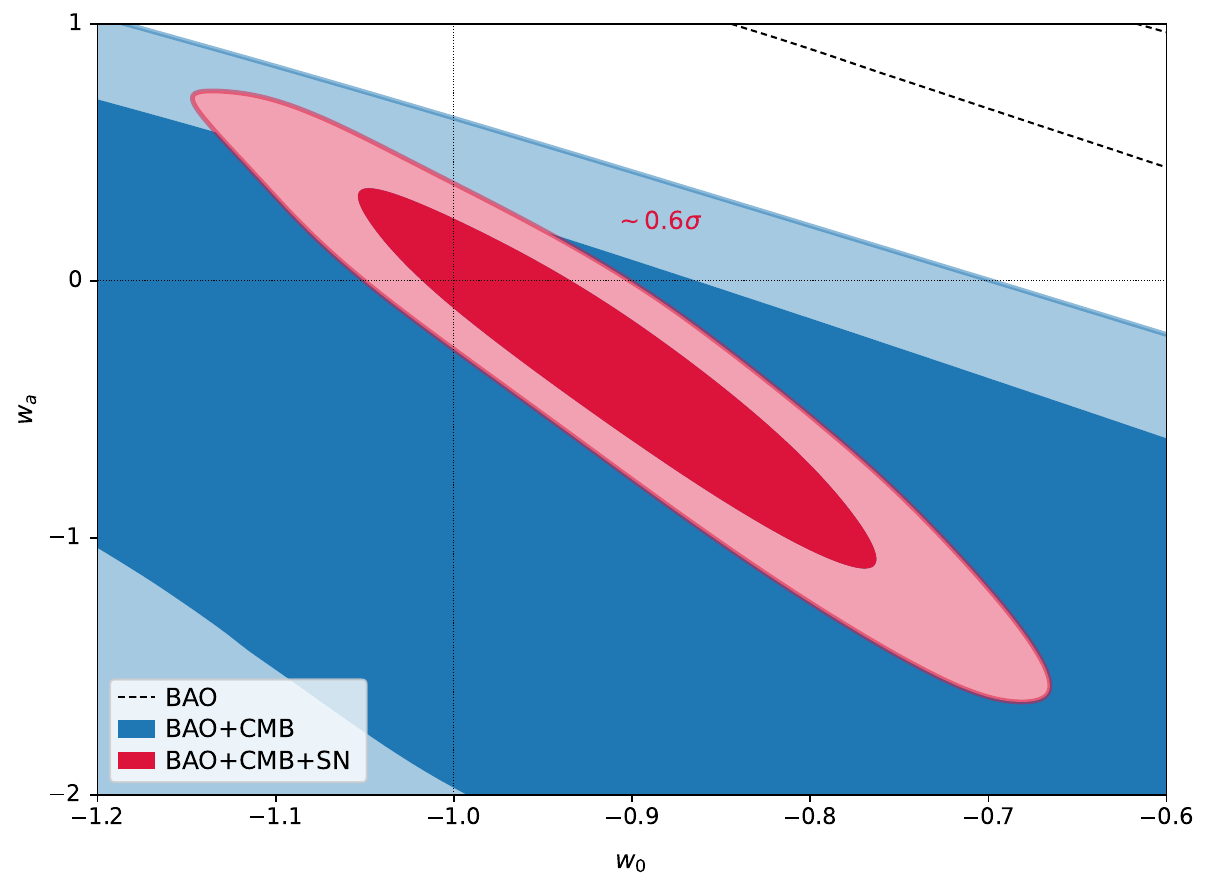}
        \caption{Redshift cut $z > 1.6$}
    \end{subfigure}
    \hfill
    \begin{subfigure}[t]{0.48\textwidth}
        \includegraphics[width=\linewidth,height=0.27\textheight]{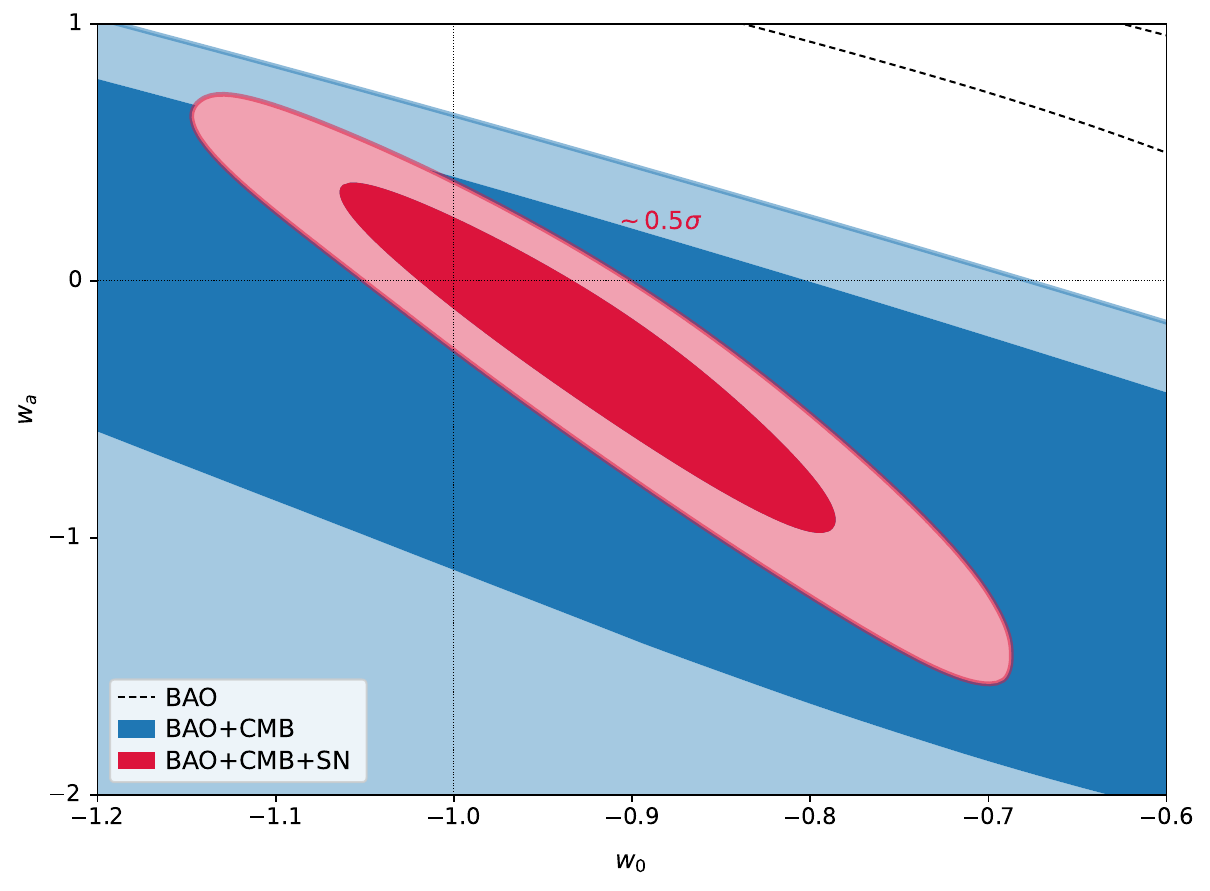}
        \caption{Redshift cut $z > 1.4$}
    \end{subfigure}
    % 每行三个子图

    \vspace{0.3cm}
    \begin{subfigure}[t]{0.48\textwidth}
        \includegraphics[width=\linewidth,height=0.27\textheight]{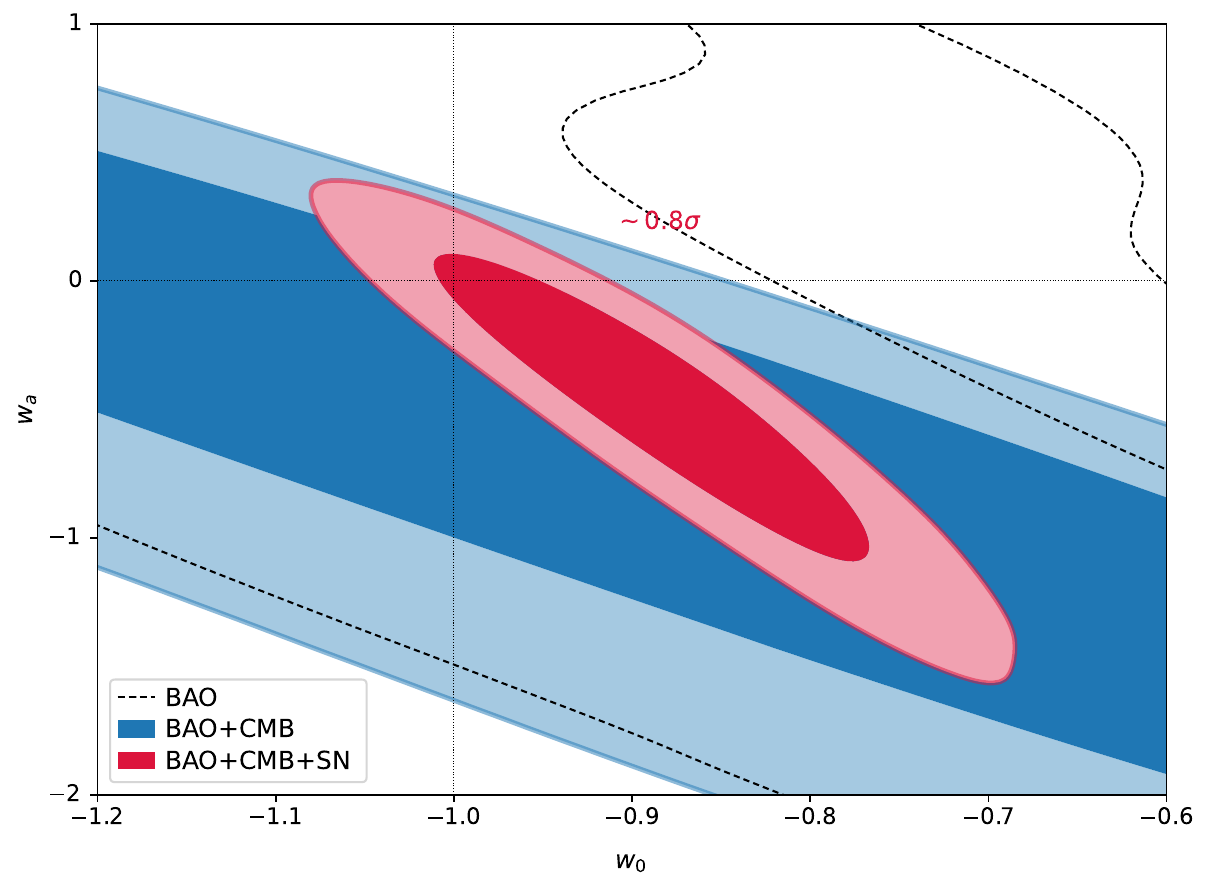}
        \caption{Redshift cut $z > 1.0$}
    \end{subfigure}
    \begin{subfigure}[t]{0.48\textwidth}
        \includegraphics[width=\linewidth,height=0.27\textheight,]{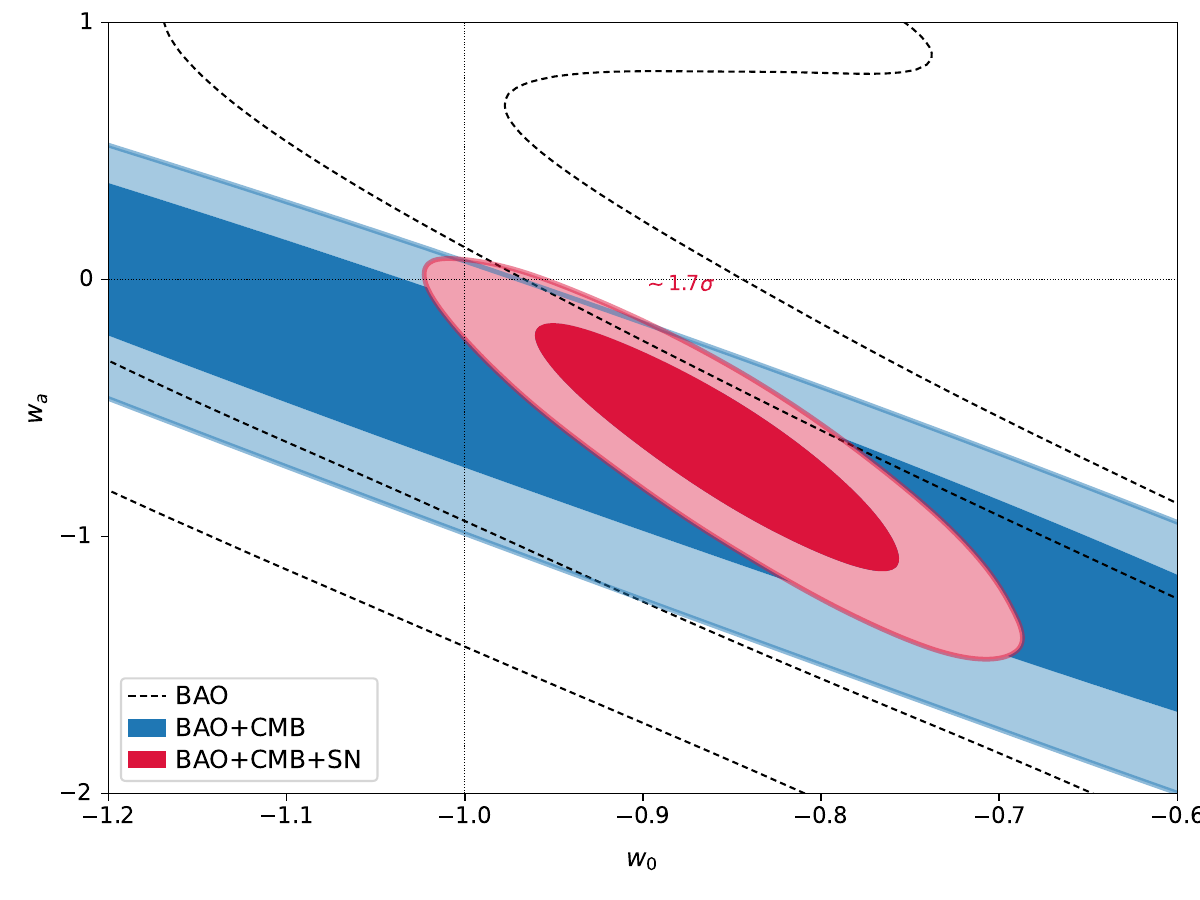}
        \caption{Redshift cut $z > 0.8$}
    \end{subfigure}
    
    \vspace{0.3cm}

    \begin{subfigure}[t]{0.48\textwidth}
        \includegraphics[width=\linewidth,height=0.27\textheight]{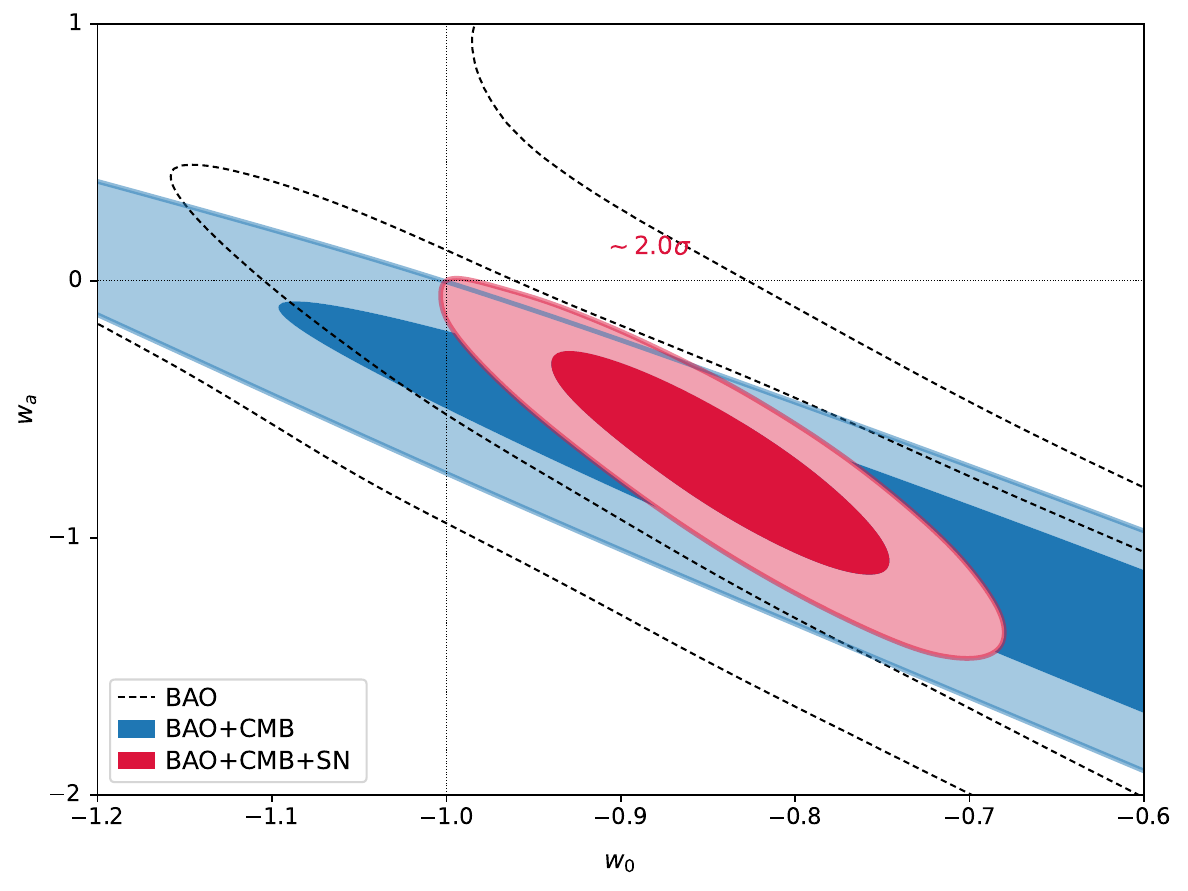}
        \caption{Redshift cut $z > 0.6$}
    \end{subfigure}
    \hfill
    \begin{subfigure}[t]{0.48\textwidth}
        \includegraphics[width=\linewidth,height=0.27\textheight]{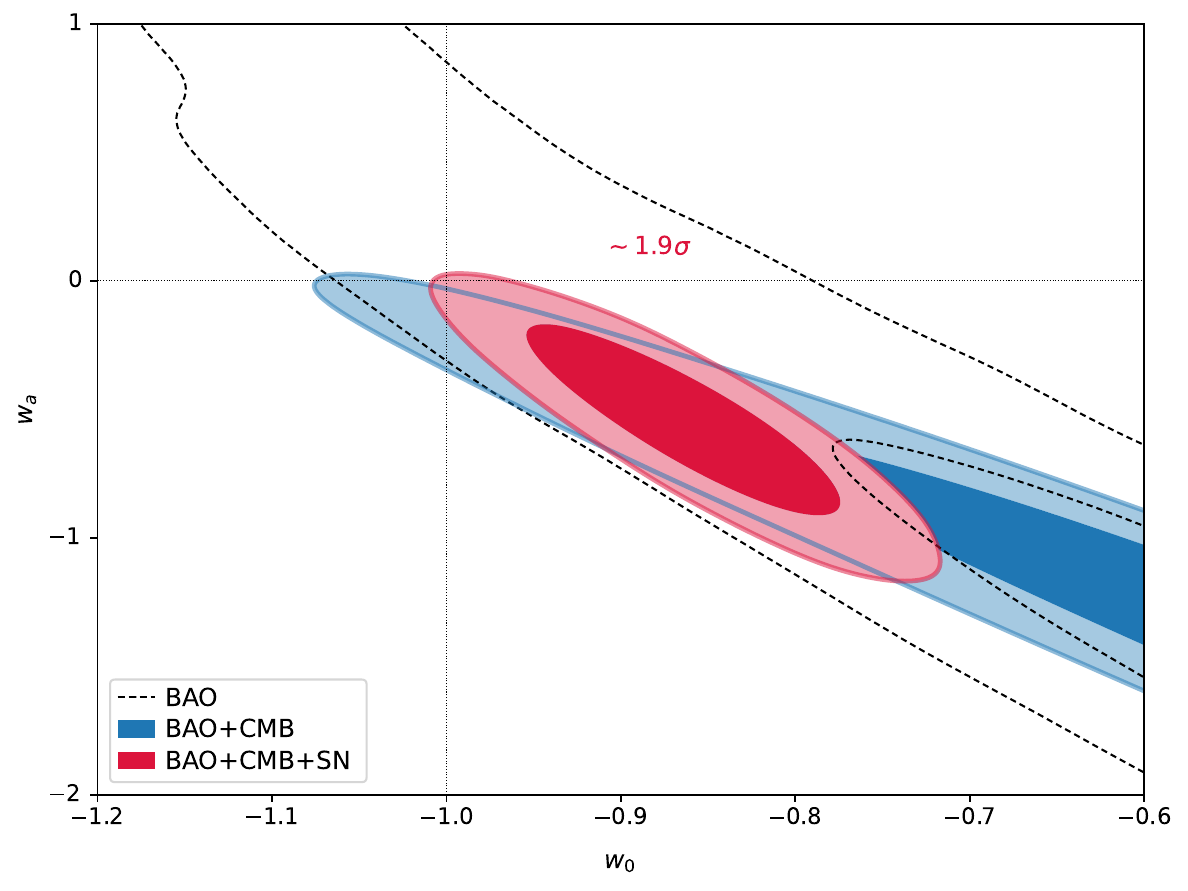}
        \caption{Redshift cut $z > 0.4$}
    \end{subfigure}

    \caption{Posterior distributions of $(w_0, w_a)$ for the $w_0w_a$CDM model obtained from DESI BAO (black dashed), BAO+CMB (blue), and BAO+CMB+SN (red) under six high-redshift path, $z>1.6$, $z>1.4$, $z>1.0$, $z>0.8$, $z>0.6$, and $z>0.4$. 
These contours indicate the 68\% and 95\% confidence levels. The gray dashed lines mark $w_0=-1$ and $w_a=0$ corresponding to the $\rm \Lambda$CDM model.
For each redshift cut, the significance of the deviation from the $\rm \Lambda$CDM model is indicated in the corresponding panel.
The deviation from the $\rm \Lambda$CDM model is quantified in each panel for the BAO+CMB+SN dataset, and we do not quote significances for BAO and BAO+CMB, as their large uncertainties.}
    \label{fig:full_page_six1}
\end{figure*}

\begin{table*}[htbp]
\centering
\caption{Summary table of cosmological parameter constraints from BAO, BAO+CMB, BAO+CMB+SN under the high-redshift path.
For each dataset and redshift selection, we report the best-fit values and their 68\% confidence levels (CLs) for the parameters ${\rm \Omega}_m$, $w_0$, $w_a$, and $H_0$ in the framework of $w_0w_a$CDM model. }
%\resizebox{\textwidth}{!}{
\begin{tabular}{ccccc}
\noalign{\smallskip}\hline\noalign{\smallskip}
\textbf{Dataset \& Cut} & ${\rm \Omega}_m$ & $w_0$ & $w_a$ & $H_0$(km/s/Mpc) \\
\noalign{\smallskip}\hline\noalign{\smallskip}
\multicolumn{5}{l}{\textbf{BAO}} \\
$z>0.4$   & $0.36^{+0.03}_{-0.04}$ & $-0.37^{+0.23}_{-0.35}$ & $-1.99^{+1.20}_{-0.72}$ & - \\
$z>0.6$   & $0.33^{+0.39}_{-0.06}$               & $-0.58^{+0.33}_{-0.47}$                & $-1.57^{+1.77}_{-1.02}$               & - \\
$z>0.8$   & $0.28^{+0.07}_{-0.06}$               & $-1.02^{+0.35}_{-0.52} $                & $-0.19^{+2.00}_{-0.94}$               & - \\
$z>1.0$   & $0.26^{+0.09}_{-0.07}$               & $-1.09^{+0.61}_{-0.56}$                & $-0.26^{+1.60}_{-1.84}$               & - \\
$z>1.4$   & $0.26^{+0.13}_{-0.09}$               & $-1.11^{+0.81}_{-0.74}$                & $-0.74^{+1.72}_{-1.57}$               & - \\
$z>1.6$   & $0.25^{+0.15}_{-0.08}$               & $-1.05^{+1.01}_{-0.76}$                & $-0.99^{+1.89}_{-1.39}$               & - \\
No redshift cut   & $0.36^{+0.03}_{-0.04}$               & $-0.44^{+0.23}_{-0.34}$                & $-1.86^{+1.24}_{-0.79}$               & - \\
\noalign{\smallskip}\hline\noalign{\smallskip}
\multicolumn{5}{l}{\textbf{BAO+CMB}} \\
$z>0.4$   & $0.36\pm0.03$ & $0.41^{+0.25}_{-0.28}$ & $-1.74^{+0.76}_{-0.71}$ & $63.20^{+2.67}_{-2.21}$ \\
$z>0.6$   & $0.33\pm0.03$               & $-0.59^{+0.32}_{-0.33}$                & $-1.43^{+0.90}_{-0.89}$               & $65.82^{+3.54}_{-3.02}$ \\
$z>0.8$   & $0.29^{+0.07}_{-0.06}$               & $-0.92^{+0.57}_{-0.59}$                & $-0.47^{+0.98}_{-1.12}$               & $69.95^{+5.28}_{-4.62}$ \\
$z>1.0$   & $0.30^{+0.07}_{-0.06}$               & $-0.92^{+0.57}_{-0.59}$                & $-0.67^{+1.32}_{-1.41}$               & $69.60^{+8.98}_{-6.89}$ \\
$z>1.4$   & $0.30^{+0.09}_{-0.07}$               & $-0.92^{+0.59}_{-0.58}$                & $-0.69\pm1.42$               & $69.02^{+10.47}_{-8.26}$ \\
$z>1.6$   & $0.29^{+0.08}_{-0.07}$               & $-0.89\pm0.52$                & $-1.06^{+1.52}_{-1.30}$               & $70.32^{+10.17}_{-8.06}$ \\
No redshift cut   & $0.35\pm0.02$               & $-0.48\pm0.24$                & $-1.59^{+0.69}_{-0.71}$               & $63.99^{+2.11}_{-1.92}$ \\
\noalign{\smallskip}\hline\noalign{\smallskip}
\multicolumn{5}{l}{\textbf{BAO+CMB+SN}} \\
$z>0.4$   & $0.31\pm0.01$ & $-0.86\pm0.06$ & $-0.55^{+0.24}_{-0.25}$ & $67.54^{+0.66}_{-0.67}$ \\
$z>0.6$   & $0.31\pm0.07$               & $-0.84\pm0.06$                & $-0.71^{+0.28}_{-0.30}$               & $68.02\pm0.71$ \\
$z>0.8$   & $0.31\pm0.01$               & $-0.86\pm0.07$                & $-0.64\pm0.32$               & $67.96^{+0.79}_{-0.76}$ \\
$z>1.0$   & $0.30^{+0.07}_{-0.06}$               & $-0.88\pm0.08$                & $-0.48^{+0.37}_{-0.42}$               & $67.61^{+0.96}_{-0.98}$ \\
$z>1.4$   & $0.32\pm0.01$               & $-0.92\pm+0.09$                & $-0.28^{+0.43}_{-0.48}$               & $67.11^{+1.13}_{-1.08}$ \\
$z>1.6$   & $0.32\pm0.01$               & $-0.90^{+0.10}_{-0.09}$                & $-0.36^{+0.44}_{-0.50}$               & $67.37^{+1.10}_{-1.23}$ \\
No redshift cut   & $0.32\pm0.01$               & $-0.87\pm0.06$                & $-0.51^{+0.22}_{-0.24}$               & $67.41^{+0.63}_{-0.61}$ \\
\noalign{\smallskip}\hline\noalign{\smallskip}
\end{tabular}
\label{tab:cosmo_params_dy}
\end{table*}

\begin{figure*}[p]  % [t] 也可以
    \centering

    % 每行三个子图
    \begin{subfigure}[t]{0.48\textwidth}
        \includegraphics[width=\linewidth,height=0.27\textheight]{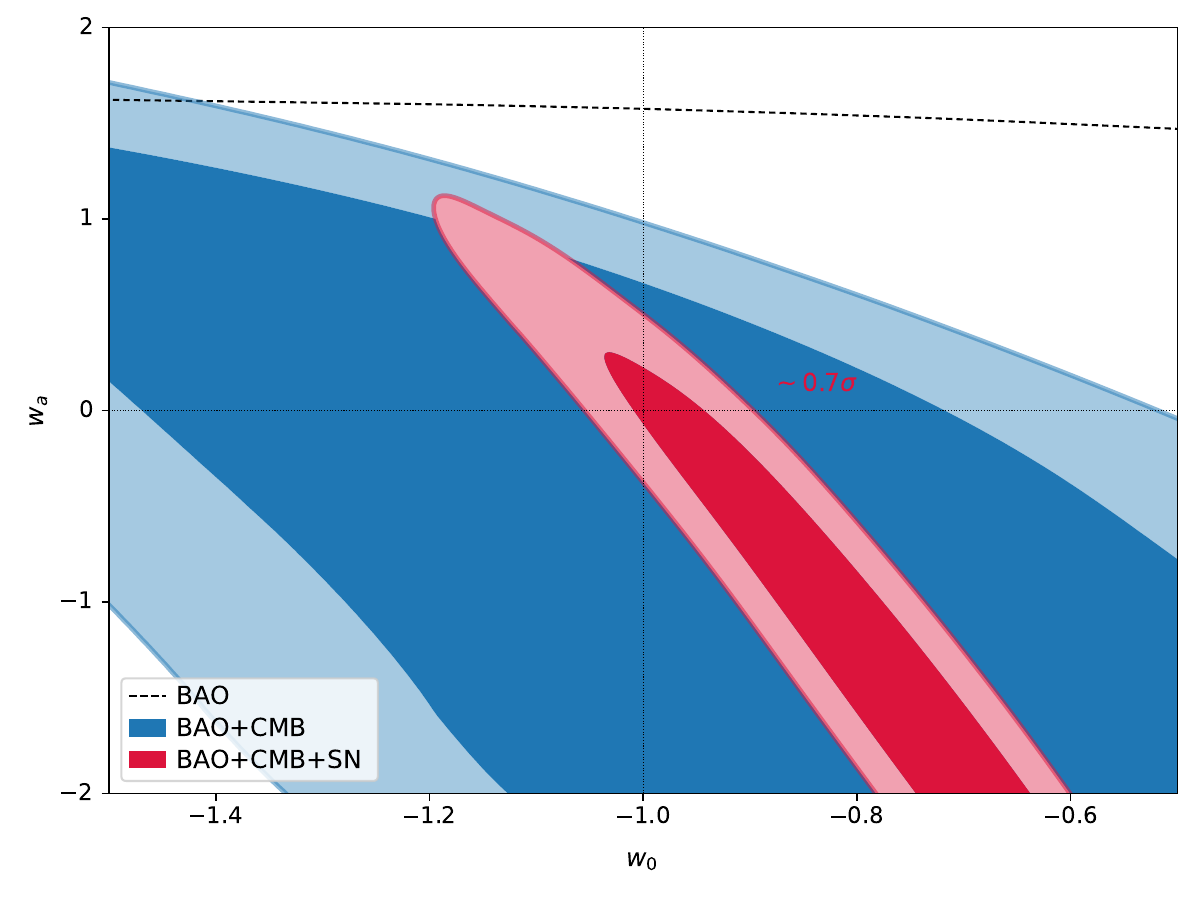}
        \caption{Redshift cut $z < 0.4$}
    \end{subfigure}
    \hfill
    \begin{subfigure}[t]{0.48\textwidth}
        \includegraphics[width=\linewidth,height=0.27\textheight]{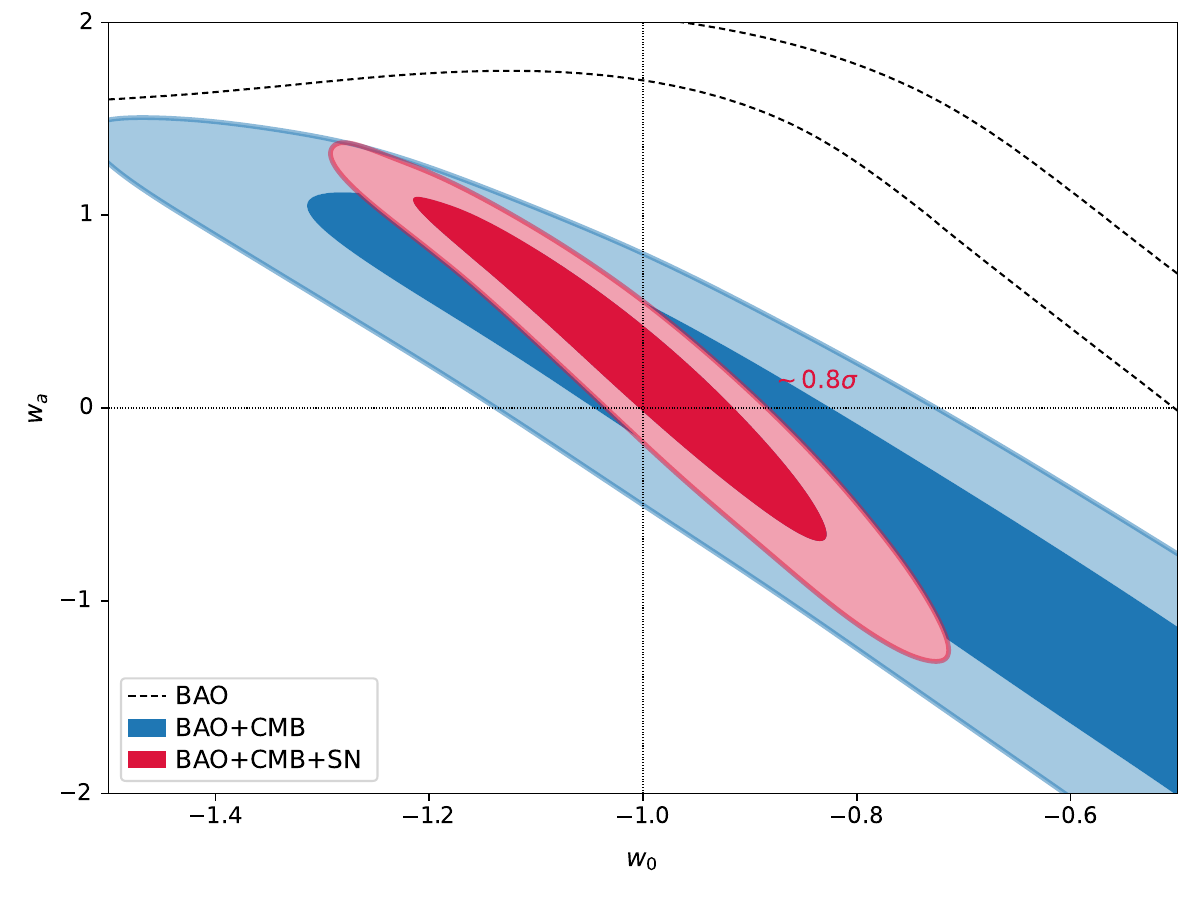}
        \caption{Redshift cut $z < 0.6$}
    \end{subfigure}

    \vspace{0.3cm}
    \begin{subfigure}[t]{0.48\textwidth}
        \includegraphics[width=\linewidth,height=0.27\textheight]{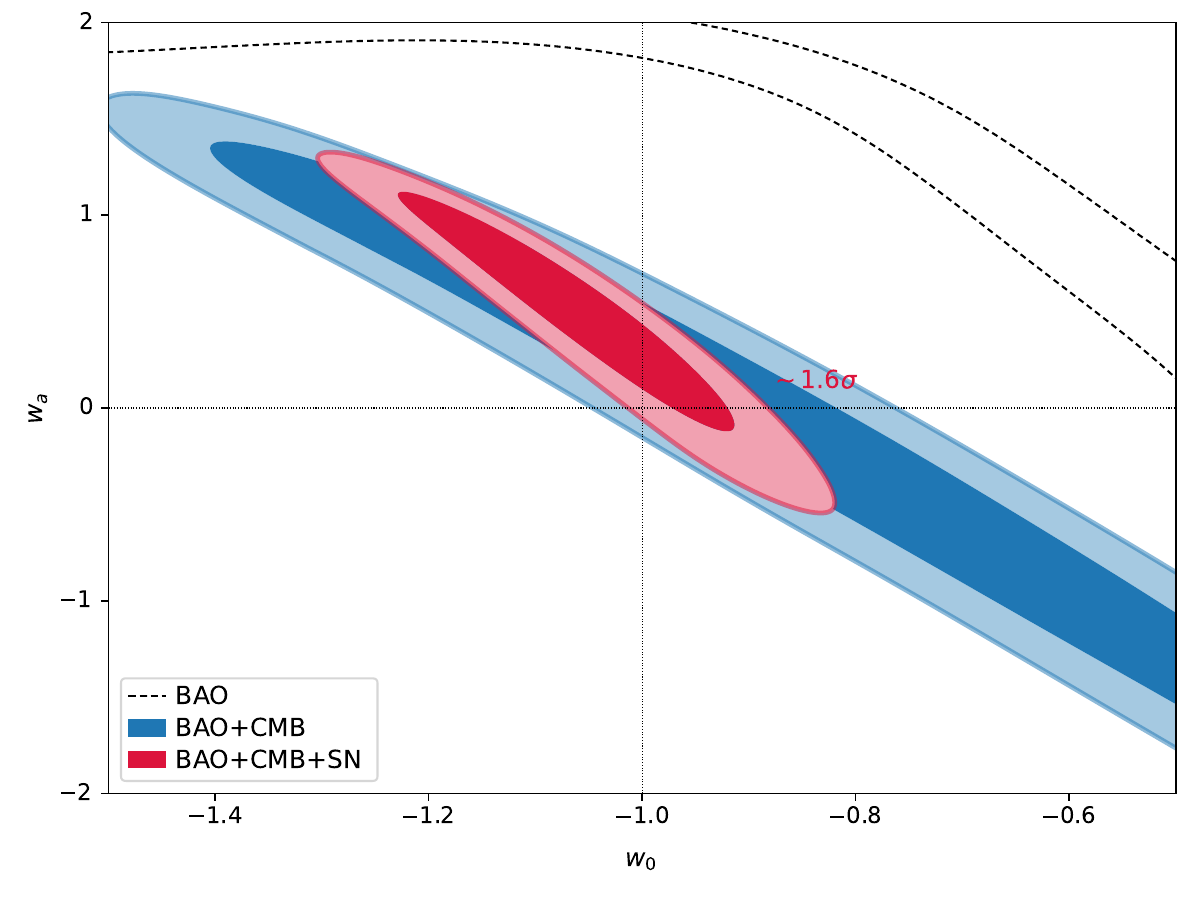}
        \caption{Redshift cut $z < 0.8$}
    \end{subfigure}
    \begin{subfigure}[t]{0.48\textwidth}
        \includegraphics[width=\linewidth,height=0.27\textheight,]{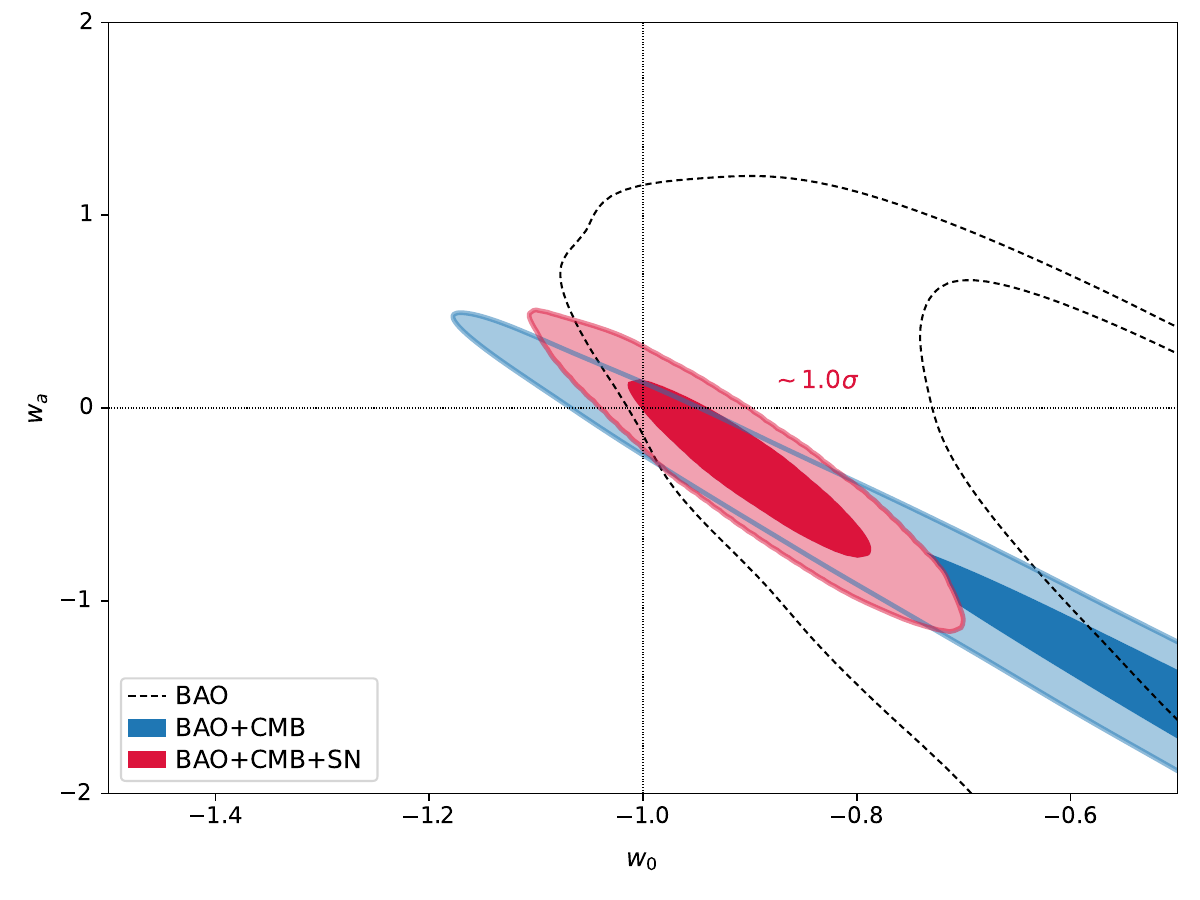}
        \caption{Redshift cut $z < 1.0$}
    \end{subfigure}
    
    \vspace{0.3cm}

    \begin{subfigure}[t]{0.48\textwidth}
        \includegraphics[width=\linewidth,height=0.27\textheight]{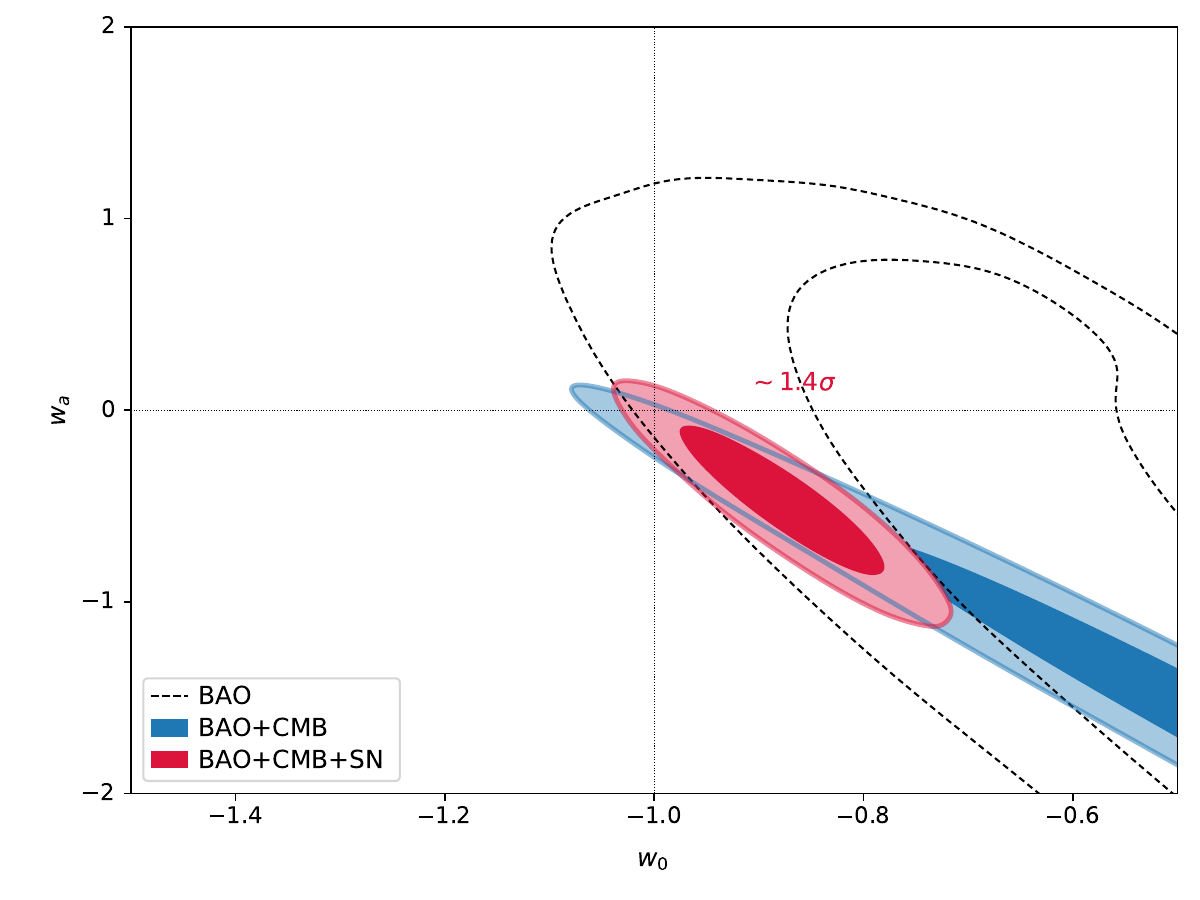}
        \caption{Redshift cut $z < 1.4$}
    \end{subfigure}
    \hfill
    \begin{subfigure}[t]{0.48\textwidth}
        \includegraphics[width=\linewidth,height=0.27\textheight]{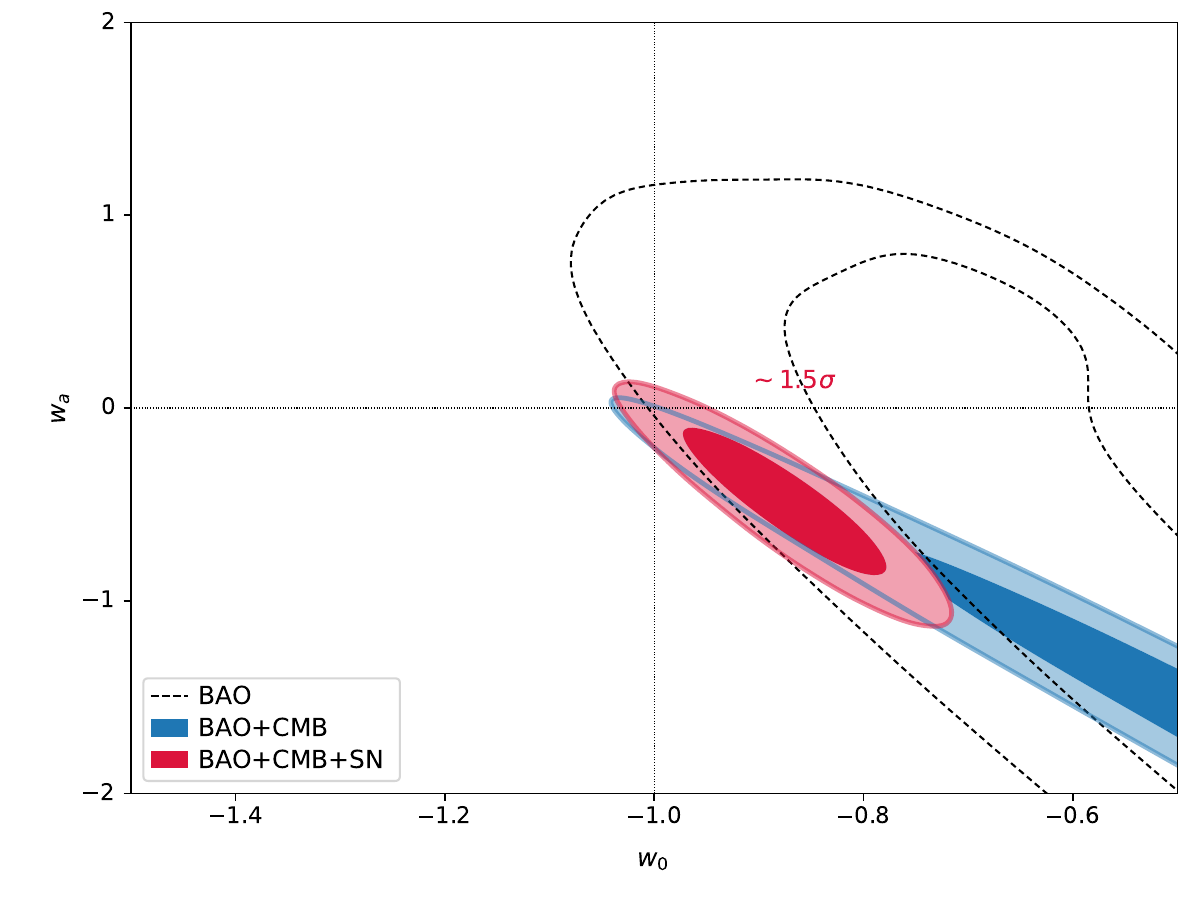}
        \caption{Redshift cut $z < 1.6$}
    \end{subfigure}

    \caption{Posterior distributions of $(w_0, w_a)$ for the $w_0w_a$CDM model obtained from DESI BAO (black dashed), BAO+CMB (blue), and BAO+CMB+SN (red) under six low-redshift path, $z<1.6$, $z<1.4$, $z<1.0$, $z<0.8$, $z<0.6$, and $z<0.4$. These contours indicate the 68\% and 95\% confidence levels. The gray dashed lines mark $w_0=-1$ and $w_a=0$ corresponding to the $\rm \Lambda$CDM model. For each redshift cut, the significance of the deviation from the $\rm \Lambda$CDM model is indicated in the corresponding panel. The deviation from the $\rm \Lambda$CDM model is quantified in each panel for the BAO+CMB+SN dataset, and we do not quote significances for BAO and BAO+CMB, as their large uncertainties.}
    \label{fig:full_page_six2}
\end{figure*}

\begin{table*}[htbp]
\centering
\caption{Summary table of cosmological parameter constraints from BAO, BAO+CMB, BAO+CMB+SN under the low-redshift path.
For each dataset and redshift selection, we report the best-fit values and their 68\% confidence levels (CLs) for the parameters ${\rm \Omega}_m$, $w_0$, $w_a$, and $H_0$ in the framework of $w_0w_a$CDM model. }
%\resizebox{\textwidth}{!}{
\begin{tabular}{lcccc}
\noalign{\smallskip}\hline\noalign{\smallskip}
\textbf{Dataset \& Cut} & ${\rm \Omega}_m$ & $w_0$ & $w_a$ & $H_0$(km/s/Mpc) \\
\noalign{\smallskip}\hline\noalign{\smallskip}
\multicolumn{5}{l}{\textbf{BAO}} \\
$z<0.4$   & $0.46^{+0.36}_{-0.31}$ & $-1.14^{+1.39}_{-1.23}$ & $-0.49^{+1.68}_{-1.71}$ & - \\
$z<0.6$   & $0.39^{+0.07}_{-0.14}$               & $-1.42^{+0.87}_{-1.04}$                & $-0.77^{+1.80}_{-1.56}$               & - \\
$z<0.8$   & $0.38^{+0.05}_{-0.08}$               & $-1.33^{+0.65}_{-0.90}$                & $-0.46^{+1.63}_{-1.71}$               & - \\
$z<1.0$   & $0.28^{+0.08}_{-0.14}$               & $-0.40^{+0.25}_{-0.28}$                & $-1.06^{+1.04}_{-1.21}$             & - \\
$z<1.4$   & $0.32^{+0.05}_{-0.11}$               & $-0.54^{+0.27}_{-0.24}$                & $-1.10^{+1.38}_{-1.30}$               & - \\
$z<1.6$   & $0.33^{+0.05}_{-0.10}$               & $-0.53^{+0.27}_{-0.24}$                & $-1.16^{+1.40}_{-1.26}$               & - \\
No redshift cut   & $0.36^{+0.03}_{-0.04}$               & $-0.44^{+0.23}_{-0.34}$                & $-1.86^{+1.24}_{-0.79}$               & - \\
\noalign{\smallskip}\hline\noalign{\smallskip}
\multicolumn{5}{l}{\textbf{BAO+CMB}} \\
$z<0.4$   & $0.29^{+0.09}_{-0.07}$ & $-0.92^{+0.51}_{-0.48}$ & $-0.94^{+1.61}_{-1.45}$ & $70.45^{+10.77}_{-9.13}$ \\
$z<0.6$   & $0.34\pm0.03$               & $-0.67^{+0.37}_{-0.42}$                & $-0.95^{+1.44}_{-1.39}$               & $64.74^{+2.80}_{-2.57}$ \\
$z<0.8$   & $0.35\pm0.02$               & $-0.77^{+0.44}_{-0.39}$                & $-0.41^{+1.18}_{-1.46}$               & $63.73^{+2.18}_{-2.05}$ \\
$z<1.0$   & $0.36^{+0.02}_{-0.03}$               & $-0.36^{+0.23}_{-0.33}$                & $-1.95^{+1.01}_{-0.71}$             & $63.49^{+2.37}_{-1.73}$ \\
$z<1.4$   & $0.35\pm0.02$               & $-0.40^{+0.24}_{-0.28}$                & $-1.82^{+0.82}_{-0.71}$               & $63.57^{+2.24}_{-1.83}$ \\
$z<1.6$   & $0.36\pm0.02$               & $-0.39^{+0.23}_{-0.27}$                & $-1.85^{+0.80}_{-0.68}$               & $63.52^{+2.23}_{-1.81}$ \\
No redshift cut   & $0.35\pm0.02$               & $-0.48\pm0.24$                & $-1.59^{+0.69}_{-0.71}$               & $63.99^{+2.11}_{-1.92}$ \\
\noalign{\smallskip}\hline\noalign{\smallskip}
\multicolumn{5}{l}{\textbf{BAO+CMB+SN}} \\
$z<0.4$   & $0.29^{+0.03}_{-0.02}$ & $-0.78^{+0.14}_{-0.18}$ & $-1.36^{+1.22}_{-1.05}$ & $70.30^{+2.55}_{-3.36}$ \\
$z<0.6$   & $0.33\pm0.02$               & $-1.00\pm0.12$                & $0.19^{+0.56}_{-0.63}$               & $65.54^{+1.77}_{-1.93}$ \\
$z<0.8$   & $0.34\pm0.01$               & $-1.05\pm0.10$                & $0.47^{+0.40}_{-0.43}$               & $64.67^{+1.25}_{-1.33}$ \\
$z<1.0$   & $0.32\pm0.01$               & $-0.90^{+0.07}_{-0.08}$                & $-0.33\pm0.30$             & $66.77\pm0.73$ \\
$z<1.4$   & $0.32\pm0.01$               & $-0.87\pm0.06$                & $-0.48^{+0.25}_{-0.26}$               & $67.35\pm0.64$ \\
$z<1.6$   & $0.32\pm0.01$               & $-0.87\pm0.06$                & $-0.49^{+0.25}_{-0.26}$               & $67.42\pm0.64$ \\
No redshift cut   & $0.32\pm0.01$               & $-0.87\pm0.06$                & $-0.51^{+0.22}_{-0.24}$               & $67.41^{+0.63}_{-0.61}$ \\
\noalign{\smallskip}\hline\noalign{\smallskip}
\end{tabular}
\label{tab:cosmo_params_xy}
\end{table*}

% Requires: \usepackage{booktabs}
% Optional but recommended: \usepackage{siunitx}
% \sisetup{round-mode=places,round-precision=2}

\begin{table*}[htbp]
\centering
\caption{{ Comparison of the $\rm \Lambda$CDM and $w_0w_a$CDM models under different redshift cuts. For each redshift cut, the minimum chi-square $\chi^2_{\min}$, the AIC and BIC values, and the corresponding differences $\Delta{\rm AIC}$ and $\Delta{\rm BIC}$ relative to the $\rm \Lambda$CDM model are presented.}
}
\begin{tabular}{lcccccccc}
\noalign{\smallskip}\hline\noalign{\smallskip}
\textbf{Redshift Cut} 
& \multicolumn{3}{c}{$\rm \Lambda$CDM} 
& \multicolumn{3}{c}{$w_0w_a$CDM}
& \multicolumn{2}{c}{$\Delta$IC} \\
\noalign{\smallskip}\hline\noalign{\smallskip}
{\textbf{$z > z_{\rm cut}$}} & $\chi^2_{\min}$ & AIC & BIC & $\chi^2_{\min}$ & AIC & BIC & $\Delta$AIC & $\Delta$BIC \\
\noalign{\smallskip}
$z>0.4$  & 235.77 & 245.77 & 264.66 & 231.69 & 245.69 & 272.13 & -0.08 & 7.48\\
\noalign{\smallskip}
$z>0.6$  & 87.55 & 97.55 & 112.33 & 83.15 & 97.15 & 117.84 &  -0.40 & 5.51\\
\noalign{\smallskip}
$z>0.8$  & 27.02 & 37.02 & 45.59 & 23.96 & 37.96 & 49.86 & 0.94 & 4.37 \\
\noalign{\smallskip}
$z>1.0$  & 16.34 & 26.34 & 33.97 & 18.70 & 32.70 & 43.38 &  6.36 & 9.41\\
\noalign{\smallskip}
$z>1.4$  & 3.35 & 13.35 & 16.89 & 3.62 & 17.62 & 22.58 & 4.27 & 5.69\\
\noalign{\smallskip}
$z>1.6$  & 1.20 & 11.20 & 12.71 & 2.24 & 16.24 & 18.36 & 5.04 & 5.65 \\

\noalign{\smallskip}\hline\noalign{\smallskip}
{\textbf{$z < z_{\rm cut}$}} & $\chi^2_{\min}$ & AIC & BIC & $\chi^2_{\min}$ & AIC & BIC & $\Delta$AIC & $\Delta$BIC \\
\noalign{\smallskip}
$z<0.4$  & 1171.96 & 1181.96 & 1208.17 & 1169.94 & 1183.94 & 1220.63 & 1.98 & 12.46\\
\noalign{\smallskip}
$z<0.6$  & 1326.52 & 1336.52 & 1363.34 & 1322.50 & 1336.50 & 1374.05 & -0.02 & 10.71\\
\noalign{\smallskip}
$z<0.8$  & 1381.83 & 1391.83 & 1418.96 & 1375.85 & 1389.85 & 1427.83 & -1.98 & 8.87\\
\noalign{\smallskip}
$z<1.0$  & 1392.72 & 1402.72 & 1429.87 & 1390.88 & 1404.88 & 1442.89 & 2.16 & 13.02\\
\noalign{\smallskip}
$z<1.4$  & 1412.27 & 1422.27 & 1449.48 & 1407.94 & 1421.94 & 1460.03 &  -0.33 & 10.55 \\
\noalign{\smallskip}
$z<1.6$  & 1414.87 & 1424.87 & 1452.09 & 1409.99 & 1423.99 & 1462.10 & -0.88 & 10.01 \\

\noalign{\smallskip}\hline\noalign{\smallskip}
\end{tabular}
\label{tab:lcdm_cpl_ic_redshift}
\end{table*}

\begin{table*}[htbp]
\centering
\caption{{ The results for parameter-shift consistency test between complementary redshift subsamples for BAO, BAO+CMB, and BAO+CMB+SN. 
For each redshift cut, the parameter-shift statistic $\chi_p^2$ and the corresponding probability-to-exceed (PTE) are listed.}}
\begin{tabular}{lcccccc}
\noalign{\smallskip}\hline\noalign{\smallskip}
\textbf{Dataset} 
& \multicolumn{2}{c}{\textbf{BAO}} 
& \multicolumn{2}{c}{\textbf{BAO+CMB}} 
& \multicolumn{2}{c}{\textbf{BAO+CMB+SN}} \\
\noalign{\smallskip}\hline\noalign{\smallskip}
  & $\chi_p^2$ & PTE & $\chi_p^2$ & PTE & $\chi_p^2$ & PTE \\
\noalign{\smallskip}
$ z_{\rm cut}=0.4$  & 0.772 & 0.680 & 0.837 & 0.658 & 1.078 & 0.583 \\
\noalign{\smallskip}
$ z_{\rm cut}=0.6$  & 0.839 & 0.657 & 0.338 & 0.844 & 1.938 & 0.379 \\
\noalign{\smallskip}
$ z_{\rm cut}=0.8$  & 0.396 & 0.820 & 3.012 & 0.222 & 5.571 & 0.062 \\
\noalign{\smallskip}
$ z_{\rm cut}=1.0$  & 1.470 & 0.479 & 0.787 & 0.675 & 0.119 & 0.942 \\
\noalign{\smallskip}
$ z_{\rm cut}=1.4$  & 0.537 & 0.764 & 0.624 & 0.732 & 0.129 & 0.937 \\
\noalign{\smallskip}
$ z_{\rm cut}=1.6$  & 0.291 & 0.864 & 0.855 & 0.652 & 0.069 & 0.966 \\

\noalign{\smallskip}\hline\noalign{\smallskip}
\end{tabular}
\label{tab:shift_consistency}
\end{table*}

In this section, we present the constraint results on $w_0$ and $w_a$ in the $w_0w_a$CDM model using DESI DR2 BAO, Pantheon+ SN Ia, and CMB distance prior data. 
We consider two complementary redshift cuts. One is the low-redshift path ($z<z_{\rm cut}$), and the other is the high-redshift path ($z>z_{\rm cut}$). 
This strategy allows us to examine how the inferred constraints on dark energy parameters vary with redshift and how different redshift ranges affect the apparent deviation from the $\rm \Lambda$CDM model.
The constraints in the $w_0$–$w_a$ plane for different redshift cuts are shown in Fig.~\ref{fig:full_page_six1}, Fig.~\ref{fig:full_page_six2}, and Fig.~\ref{fig:all}. 
The best-fit values and the corresponding 68\% confidence intervals of the $w_0w_a$CDM parameters are listed in Table~\ref{tab:cosmo_params_dy} and Table~\ref{tab:cosmo_params_xy}.

We first examine the high-redshift path ($z>z_{\rm cut}$), shown in Fig.~\ref{fig:full_page_six1}. 
In this case, the $\rm \Lambda$CDM model $\{w_0,w_a\}=\{-1,0\}$ always remains inside the $95\%$ confidence level region. 
The constraints from only DESI BAO data are broad with large uncertainties, while the inclusion of CMB data significantly tighten the constraints on $\{w_0,w_a\}$. 
The addition of SNe Ia further breaks the degeneracy between $w_0$ and $w_a$ and provides the tightest constraints.
From the BAO+CMB+SN combination we obtain $\{w_0,w_a\}=\{-0.88\pm0.08,-0.48^{+0.37}_{-0.42}\}$ at $z>1.0$, $\{w_0,w_a\}=\{-0.92\pm0.09,-0.28^{+0.43}_{-0.48}\}$ at $z>1.4$, and $\{w_0,w_a\}=\{-0.90^{+0.10}_{-0.09},-0.36^{+0.44}_{-0.50}\}$ at $z>1.6$. 
These results suggest that excluding low-redshift data points weakens the apparent preference for dynamical dark energy, partly due to the reduced constraining power when fewer data points are included.
As seen in Fig.~\ref{fig:full_page_six1}, larger shifts away from the $\rm \Lambda$CDM model are observed in the cases of $z>0.6$ and $z>0.4$, reaching a level of $\sim2\sigma$. This behavior is consistent with previous studies~\cite{LRG12_ratio_or_monopole,Zheng:2024qzi} that have reported similar shifts when including specific DESI LRG measurements. However, it should be noted that removing low-redshift data points reduces the statistical constraining power of the dataset, which may partly account for the apparent significance of these shifts.

Then, we turn to the low-redshift path ($z<z_{\rm cut}$). 
In this case, the constraints also remain consistent with $\rm \Lambda$CDM at the $95\%$ confidence level, but the evolution with $z_{\rm cut}$ provides additional insight. 
A similar trend appears in the low-redshift path. 
The constraints from BAO measurements remain broad because they provide relatively weak constraints on the parameters $w_0$ and $w_a$. 
Including CMB data reduces the uncertainties, while the addition of SNe Ia significantly tightens the constraints.
Using the low-redshift BAO+CMB+SN dataset ($z<0.4$), we obtain 
$\{w_0,w_a\}=\{-0.78^{+0.14}_{-0.18},-1.36^{+1.21}_{-1.04}\}$ 
with $H_0=70.30^{+2.55}_{-3.36}~{\rm km\,s^{-1}\,Mpc^{-1}}$. 
At this stage, the constraints remain relatively weak and are fully consistent with the $\rm \Lambda$CDM model within the $95\%$ confidence level.
As higher-redshift data are gradually included, larger shifts away from the $\rm \Lambda$CDM model are observed. In particular, when the dataset is extended to $z<0.8$, the shift reaches approximately $1.5\sigma$. This redshift range coincides with the inclusion of the DESI BAO LRG1 ($z_{\rm eff}\approx0.51$) and LRG2 ($z_{\rm eff}\approx0.71$) measurements, which have been associated in previous studies with similar shifts in the inferred cosmological constraints.
When even higher-redshift data are incorporated (e.g. $z<1.4$ and $z<1.6$), the overall constraints become tighter due to the increased number of datapoints. 
However, the significance of the deviation does not increase substantially and remains at roughly the $\sim1.5\sigma$ level. 
This behavior indicates that although additional high-redshift data improve the statistical precision of the constraints, they do not significantly strengthen the preference for dynamical dark energy.

%{ From the above results, the two redshift-cut strategies lead to a consistent picture. First, the BAO measurements alone provide relatively weak constraints and the inclusion of CMB information significantly reduces the uncertainties. The addition of SNe Ia further breaks the degeneracy between $w_0$ and $w_a$ and provides the tightest constraints. Second, the apparent deviation from the $\Lambda$CDM model is primarily associated with intermediate redshift data around $z\sim0.4$–$0.8$. This redshift range includes the DESI BAO LRG1 and LRG2 BAO measurements, which has been discussed in previous studies in relation to the deviation from the $\Lambda$CDM model. Finally, we find that most of the shift is driven by changes in $w_a$, while $w_0$ remains close to $-1$ across the different redshift selections. However, in all cases, the $\Lambda$CDM model remains within the $95\%$ confidence level.}

{To quantitatively assess the model preference under different redshift-cut subsamples, we introduce the Akaike Information Criterion (AIC) and the Bayesian Information Criterion (BIC) to compare the $w_0w_a$CDM model with the $\rm \Lambda$CDM model. 
They are defined as
\begin{equation}
\mathrm{AIC} = \chi^2_{\min} + 2k ,
\end{equation}
\begin{equation}
\mathrm{BIC} = \chi^2_{\min} + k \ln N ,
\end{equation}
where $\chi^2_{\min}$ is the minimum chi-square value obtained from the best-fit values of the model, $k$ denotes the number of free parameters, and $N$ is the number of data points retained after applying a given redshift cut. 
For each redshift-cut subsample, we compute $\Delta$AIC and $\Delta$BIC relative to the $\rm \Lambda$CDM model to evaluate the relative statistical preference between the two models. It should be noted that $\Delta\mathrm{AIC}\gtrsim 2$ indicates mild evidence, while $\Delta\mathrm{AIC}\gtrsim 6$ suggests a strong preference for the model with a smaller AIC. Because the BAO and BAO+CMB datasets contain relatively few datapoints and provide weak constraints on the $w_0w_a$CDM model, we focus on the BAO+CMB+SN combination when evaluating the information criteria. The model comparison results based on the information criteria are summarized in Table~\ref{tab:lcdm_cpl_ic_redshift}. 
We find that the relative preference between the $\rm \Lambda$CDM and $w_0w_a$CDM models does not change significantly under different redshift cuts.
From the perspective of the Akaike Information Criterion, the values of $\Delta$AIC are generally small. 
For example, in the high-redshift path, the cases $z>0.4$, $z>0.6$, and $z>0.8$ all give $|\Delta{\rm AIC}|<1$, indicating that the current data do not provide statistically significant evidence favoring either model. 
Even $\Delta$AIC reaches values of order $\sim4$--$6$, the evidence remains at most weak and does not indicate a decisive preference for the dynamical dark energy model.
In contrast, the Bayesian Information Criterion consistently favors the $\rm \Lambda$CDM model in all cases. 
Since BIC offers a stronger penalty for additional model parameters, the larger values of $\Delta$BIC indicate that the extra degrees of freedom in the $w_0w_a$CDM model are not sufficiently justified by the improvement in the fit.
Overall, these results suggest that varying the redshift cut does not lead to a robust change in the model preference. 
While the parameter contours may shift under different redshift-cut subsamples, the information criteria indicate that the current data are not sufficient to establish a statistically significant preference for dynamical dark energy over the $\rm \Lambda$CDM model.}

{ Following the Ref.~\cite{planck2018}, we assess the consistency of the parameter differences between complementary redshift subsamples ($z<z_{\rm cut}$ versus $z>z_{\rm cut}$) in the $(w_0,w_a)$ subspace. For two independent datasets (or complementary redshift subsamples) $A$ and $B$, the parameter difference vector is defined as
\begin{equation}
\Delta \mathbf{p} = \boldsymbol{\mu}_A - \boldsymbol{\mu}_B ,
\end{equation}
where $\boldsymbol{\mu}_A$ and $\boldsymbol{\mu}_B$ denote the posterior mean parameter vectors inferred from the respective datasets. 
Assuming statistical independence, the covariance matrix of the parameter difference is
\begin{equation}
\mathbf{C}_{\Delta} = \mathbf{C}_A + \mathbf{C}_B ,
\end{equation}
where $\mathbf{C}_A$ and $\mathbf{C}_B$ are the parameter covariance matrices estimated from the corresponding MCMC chains. 
The parameter-shift statistic is then defined as
\begin{equation}
\chi_p^2 = \Delta \mathbf{p}^{\rm T} \mathbf{C}_{\Delta}^{-1} \Delta \mathbf{p}.
\end{equation}
Under the null hypothesis that the two subsamples are statistically consistent realizations of the same underlying cosmology, $\chi_p^2$ follows a $\chi^2$ distribution with degrees of freedom equal to the dimension of $\mathbf{p}$. 
The level of consistency is quantified by the probability-to-exceed (PTE),
\begin{equation}
\mathrm{PTE} = \mathrm{Prob}\left(\chi^2 \ge \chi_p^2\right).
\end{equation}
Assuming an approximately multivariate Gaussian posterior distribution for the parameters, we compute the PTE from the $\chi^2$ distribution with $\nu=2$ to assess the statistical consistency between the two complementary subsamples. The results are summarized in Table~\ref{tab:shift_consistency}. For BAO and BAO+CMB, all redshift cuts give large PTE values. 
For example, ${\rm PTE}\simeq0.48$--$0.86$ for BAO and ${\rm PTE}\simeq0.22$--$0.84$ for BAO+CMB. 
These results indicate that there is no statistically significant inconsistency between the complementary subsamples.
For BAO+CMB+SN, the PTE values are generally close to unity for $z_{\rm cut}\ge1.0$, e.g. ${\rm PTE}=0.942$, $0.937$, and $0.966$ for $z_{\rm cut}=1.0$, $1.4$, and $1.6$, respectively. These results show that the cosmological constraints derived from the two complementary subsamples are consistent with each other for $z_{\rm cut}=1.0$, $1.4$, and $1.6$.
The largest shift occurs at $z_{\rm cut}=0.8$ for BAO+CMB+SN, where $\chi_p^2=5.571$ with ${\rm PTE}=0.062$. 
Nevertheless, this value does not provide statistically significant evidence for inconsistency and can be interpreted as a mild statistical fluctuation.}

A number of recent studies have examined the reported deviations from the $\rm \Lambda$CDM model using different combinations of cosmological datasets. 
For example, Ref.~\cite{Binned_w_DESI} reconstructed the dark-energy equation of state using three redshift bins in $w_{\mathrm{bin}}(z)$ based on DESI BAO combined with CMB and several SN samples. 
Their results indicate mild low-redshift shifts ($w>-1$ at $\sim1$–$3\sigma$) and weak constraints at higher redshift, suggesting that a sharp phantom–to–quintessence transition is not conclusively established. 
Ref.~\cite{LRG12_ratio_or_monopole} further showed that similar shifts may be associated with the DESI LRG1 and LRG2 BAO measurements. 
Similarly, Ref.~\cite{BAO_SN_tomography} compared luminosity distances derived from BAO and the Pantheonplus SN sample using a redshift-tomography approach and found that the discrepancy is more pronounced at low redshift ($0.1 \lesssim z \lesssim 0.8$), while becoming weaker at higher redshift ($0.8 \lesssim z \lesssim 2.3$). 
In addition, Ref.~\cite{Huang:2025som} investigated the DESI DR1/DR2 evidence for dynamical dark energy and argued that the apparent preference may be largely influenced by the treatment of low-redshift supernovae in the DESY5 compilation.
These studies suggest that the reported deviation from the $\rm \Lambda$CDM model may be sensitive to the properties of low-redshift data and to dataset selection. 
However, most previous analyses have focused on specific subsamples or particular dataset combinations. 
In contrast, our complementary redshift-cut analysis provides a systematic way to examine how the inferred constraints on dark energy parameters vary across redshift by comparing two complementary subsamples ($z<z_{\rm cut}$ and $z>z_{\rm cut}$). 
This approach allows us to assess the stability of the apparent deviation and to quantitatively test the statistical consistency between complementary redshift subsamples. 
We find that the BAO and SN Ia data in the redshift interval $z\sim0.4$–$0.8$ are associated with larger shifts in the inferred constraints.
Compared with previous analyses based on specific subsample selections, our analysis provides a more systematic and statistically consistent perspective for assessing the robustness of the apparent deviation.

\section{Conclusions}
\label{sec:con}

Recent cosmological analyses from DESI DR1 and DR2 have attracted considerable attention and sparked debate regarding potential challenges to the standard $\rm \Lambda$CDM model.
In this work, we investigate how the inferred constraints vary across different redshift selections within the $w_0w_a$CDM model and within the fixed framework of DESI BAO DR2, Pantheon+, and compressed CMB distance priors.
We apply complementary redshift cuts and combine information-criterion model comparison with a parameter-shift consistency test to quantify the dependence of the inferred constraints on redshift selection and to assess the statistical consistency between complementary subsamples.

Our main findings can be summarized as follows.
First, in all cases considered here, the $\rm \Lambda$CDM model remains consistent within the $95\%$ confidence level.
This result should be interpreted within the compressed CMB distance-prior framework adopted in this work, which is not equivalent to the full CMB power-spectrum likelihood and may lead to different levels of model preference and parameter consistency.
Second, the redshift-cut analysis shows that the apparent parameter shifts are most pronounced when data in the redshift range $z\sim0.4$--$0.8$ are included, while the inclusion of higher-redshift measurements generally brings the results closer to the $\rm \Lambda$CDM expectation.
Third, within this data framework, the information-criterion analysis shows no statistically significant preference between the $w_0w_a$CDM and $\rm \Lambda$CDM models according to the AIC, while the BIC consistently favors the simpler $\rm \Lambda$CDM model.
Finally, the parameter-shift test shows no statistically significant tension between the complementary redshift subsamples, as all PTE values remain above $0.05$.

Within the DESI BAO DR2, Pantheon+, and compressed CMB distance-prior framework adopted here, these results do not provide statistically robust evidence favoring the $w_0w_a$CDM model over $\rm \Lambda$CDM.
The observed shifts may reflect the sensitivity of the inferred constraints to redshift selection, statistical fluctuations, and the limited constraining power of the current subsamples. We also emphasize that, since this work uses Pantheon+ as the only SNe Ia compilation, the conclusions should not be generalized to other SNe Ia samples without further analysis.

Looking ahead, future observations will be essential for clarifying the origin of the deviations discussed above. 
The redshift cut strategy adopted in this work can be further refined with upcoming DESI data releases and next-generation SNe Ia samples. 
In particular, improved low-redshift spectroscopy and better control of SNe Ia systematics will be crucial for determining whether the observed deviations arise from observational systematics or hint at possible new physics. 
Moreover, the inclusion of other independent cosmological probes, such as strong gravitational lensing (SGL) and gravitational waves (GWs), will provide complementary constraints on the nature of dark energy.

\section*{Acknowledgements}
This work is supported by the National Natural Science Foundation of China under Grant No. 12403002, 12505070, 12305059, and 12433001; The Henan Provincial Natural Science Foundation No. 252300420902; The Joint Fund of Henan Province Science and Technology R\&D Program No.235200810111; The Startup Research Fund of Henan Academy of Sciences No. 241841221, 241841222, 241841224; The Scientific and Technological Research Project of Henan Academy of Science No. 20252345001, 20252345003; The Henan Province High-Level Talent Internationalization Cultivation Program No. 12024032.

\section*{DATA AVAILABILITY STATEMENTS}

The data underlying this article will be shared on reasonable request
to the corresponding author.

%\begin{acknowledgements}
%If you'd like to thank anyone, place your comments here
%and remove the percent signs.
%\end{acknowledgements}
  
% BibTeX users please use one of
%\bibliographystyle{spbasic}      % basic style, author-year citations

%\bibliographystyle{spmpsci}      % mathematics and physical sciences
\bibliographystyle{spphys}       % APS-like style for physics
\bibliography{refer}   % name your BibTeX data base

% Non-BibTeX users please use
\end{document}